\documentclass[iop]{emulateapj}
\usepackage{graphicx}
\usepackage{epsfig}
\usepackage{amsmath,amssymb}
\usepackage{verbatim}
\usepackage{rotating}
\usepackage{subfigure}
\usepackage{epstopdf}
\usepackage[hang,flushmargin]{footmisc} 

\def\lapprox{\lower.4ex\hbox{$\;\buildrel <\over{\scriptstyle\sim}\;$}}
\def\gapprox{\lower.4ex\hbox{$\;\buildrel >\over{\scriptstyle\sim}\;$}}

\shorttitle{ROSAT Stacking}
\shortauthors{Anderson et al.}

\begin{document}
\title{Extended Hot Halos Around Isolated Galaxies Observed in the \emph{ROSAT} All-Sky Survey}
\author{Michael E. Anderson\altaffilmark{1}, Joel N. Bregman\altaffilmark{1}, Xinyu Dai\altaffilmark{2}}
\altaffiltext{1}{Department of Astronomy, University of Michigan, Ann Arbor, MI 48109; 
michevan@umich.edu, jbregman@umich.edu}
\altaffiltext{2}{Homer L. Dodge Department of Physics and Astronomy, University of Oklahoma, Norman, OK 73019, xdai@ou.edu}

\begin{abstract}
We place general constraints on the luminosity and mass of hot X-ray emitting gas residing in extended ``hot halos'' around nearby massive galaxies. We examine stacked images of 2165 galaxies from the 2MASS Isolated Galaxy Catalog (2MIG), as well as subsets of this sample based on galaxy morphology and K-band luminosity. We detect X-ray emission at high confidence (ranging up to nearly $10\sigma$) for each subsample of galaxies. The average $L_X$ within 50 kpc is $1.0\pm0.1$ (statistical) $\pm0.2$ (systematic) $\times10^{40}$ erg s$^{-1}$, although the early-type galaxies are more than twice as luminous as the late-type galaxies. Using a spatial analysis, we also find evidence for extended emission around five out of seven subsamples (the full sample, the luminous galaxies, early-type galaxies, luminous late-type galaxies, and luminous early-type galaxies) at 92.7\%, 99.3\%, 89.3\%, 98.7\%, and 92.1\% confidence, respectively. Several additional lines of evidence also support this conclusion and suggest that about 1/2 of the total emission is extended, and about 1/3 of the extended emission comes from hot gas.  For the sample of luminous galaxies, which has the strongest evidence for extended emission, the average hot gas mass is $4\times10^9 M_{\odot}$ within 50 kpc and the implied accretion rate is $0.4 M_{\odot}$ yr$^{-1}$.
 
\end{abstract}
 

\maketitle

\section{Introduction}

In addition to the central black hole, the stellar component (bulge and/or disk) and stellar halo, the dark matter halo, and interstellar gas, a fundamental component of an isolated massive galaxy is the circumgalactic medium. The CGM contains all the material outflowing from or accreting onto the galaxy, and may contain a quasi-static hot gaseous halo as well. As the medium through which baryons flow into and out of a galaxy, the CGM therefore encodes a wealth of information about the galaxy's past and future. Understanding the CGM is critical for understanding the baryonic processes responsible for galaxy formation. A portion of the CGM gas has been observed extensively by studying UV absorption lines, at redshifts from 0 to $\sim 2$, probing temperatures from $10^4$ K to $10^{5.8}$ K (e.g. \citealt{Tripp2000}, \citealt{Adelberger2003}, \citealt{Tumlinson2011}, \citealt{Rudie2012}) . But a significant fraction is expected to lie at higher temperatures, in the X-ray regime (see next paragraph).  This hot component of the CGM is the focus of this paper. 

The hot gas is generally assumed to be associated with a quasi-static hot gaseous halo \citep{White1978}, although outflows can also reach the required temperatures \citep{Strickland2000}, and hot filamentary accretion may be possible as well \citep{Sijaki2012}. Present-day simulations predict that hot halos form as gas accretes onto a galaxy; if the dark matter halo is sufficiently massive ($\sim 10^{12} M_{\odot}$), some of the inflowing gas is heated by an accretion shock to approximately the virial temperature of the system (around a quarter of a keV for an $L*$ galaxy; \citealt{Birnboim2003}). Below the critical galaxy mass, the majority of the gas is thought to accrete in a ``cold flow''. 

The hot gaseous halo is thought to be responsible for a number of observed effects. It makes accretion onto a galaxy much slower, strangling star formation and potentially explaining the color-magnitude bimodality \citep{Dekel2006} as an artifact of the transition between the cold and hot modes of accretion. The increasing prevalence of hot mode accretion at lower redshift also can help to explain the observed ``down-sizing'' in the star formation history (\citealt{Bower2006}, \citealt{DeLucia2006}). For the very largest galaxies, with inferred dark matter halos as massive as small groups, there is  some conceptual overlap between the hot halo and the intergroup medium (as in the case of the so-called ``fossil groups'', e.g. \citealt{Mulchaey1999}).

Unfortunately, there is scant observational evidence even for the existence of extended gaseous halos caused by hot mode accretion onto large galaxies. This is especially true for late-type galaxies. Searches for hot halo emission around late-type galaxies were first conducted with the Einstein Observatory (e.g. \citealt{Bregman1982}, \citealt{McCammon1984}) and then later with \emph{ROSAT} (e.g. \citealt{Benson2000}); these studies were only able to place upper limits on the extended diffuse emission. With the newer generation of X-ray telescopes, diffuse emission has been detected around late-type galaxies (e.g. \citealt{Strickland2004a}, \citealt{Li2006}, \citealt{Tullmann2006}, \citealt{Owen2009}, \citealt{Yamasaki2009}). This emission is generally confined to the innermost 10-20 kpc and is seen perpendicular to the disk, and widely assumed to stem from internal processes such as galactic fountains. Hot gas is also observed around the Milky Way (e.g. \citealt{Snowden1998}), but its radial extent is unclear (\citealt{Anderson2010}, \citealt{Gupta2012}, Miller and Bregman 2012). So far, the only spiral galaxies with strong evidence for genuinely extended hot gaseous halos are NGC 1961 and UGC 12951, two of the largest spiral galaxies in the Universe (\citealt{Anderson2011}; \citealt{Dai2011}). 

The situation around early-type galaxies is rather different, since the Einstein Observatory was able to detect hot halos around massive ellipticals (\citealt{Forman1979}; \citealt{Forman1985}). These hot halos vary by more than an order of magnitude in their X-ray luminosity for a fixed stellar luminosity, and studying this scaling relation continues to be an important area of research today (\citealt{O'Sullivan2001}, \citealt{Mulchaey2010}, \citealt{Bogdan2012}). However, the hot halo metallicities are generally super-solar \citep{Mathews2003} and so the gas is generally assumed to have an internal origin (i.e., AGB stellar winds shock-heated to the virial temperature) instead of revealing hot mode accretion, although it is possible that the high-metallicity gas is screening a lower-metallicity component \citep{Crain2010}. 

There are also significant theoretical uncertainties about the amount of mass contained in hot gaseous halos. Analytic calculations (\citealt{White1991}; \citealt{Fukugita2006}) typically assume that the hot halo contains enough baryonic mass to bring the baryon fraction within the virial radius up to the cosmic mean of 0.17 \citep{Dunkley2009}. In simulations, the baryon fraction of a galaxy can vary substantially, depending on the initial conditions and the feedback prescription. To illustrate this, we can take the example of the Milky Way. The stars and ISM gas of the Milky Way only contribute about 20-30\% of the baryons required to bring our Galaxy's baryon fraction to the cosmic mean, and these quantities can be reproduced in simulations. But different simulations predict very different values for the total baryon fraction of the Galaxy, and these differences are nearly entirely due to the mass of the hot halo. As examples, \citet{Guedes2011} ran a self-consistent simulation of a Milky Way-like galaxy and found a total baryon fraction of 70\% of the cosmic value, while \citet{Moster2011} ran simulations of Milky Way-like galaxies initialized to 50\% and 100\% of the cosmic value. Finally, \citet{Faucher-Giguere2011} compared various feedback  prescriptions and were able to produce $L*$ galaxies with baryon fractions ranging from 30\% (essentially no hot halo) to 100\% of the cosmic value. Modern simulations of galaxy formation use baryon fractions all over this range, such that the baryon fraction of galaxies is essentially a free parameter.

Unfortunately, observations have not been sufficient to help constrain the mass of hot halos around $L*$ galaxies. We collected various observational constraints in \citet{Anderson2010} which collectively suggest that hot halos do not contain more than 10--20\% of the ``missing'' baryons from galaxies. We also detected extended X-ray emission around two giant spiral galaxies, NGC 1961 \citep{Anderson2011} and UGC 12591 \citep{Dai2011}, and interpreted the emission as hot halo emission; these galaxies only seem to contain about 10-20\% of the missing baryons in their hot halos as well. On the other hand, \citet{Humphrey2011} detected a hot halo around the isolated $L*$ elliptical NGC 720, and that hot halo seems to bring the baryon fraction of the galaxy up to the cosmic mean. The recent observation of NGC 1521 \citep{Humphrey2012} seems to find a similar hot halo around this isolated elliptical as well. The discrepancy between the inferred (low) baryon fractions in our giant spirals and the (high) baryon fractions in the L* ellipticals of Humphrey et al. merits further study, which we will undertake in subsequent work. 

With only a few individual cases, and the aforementioned discrepancy in the results, it has not yet been possible to draw broader conclusions about the mass contained within hot halos. In this paper, we examine the aggregate properties of thousands of galaxies, by stacking images from the \emph{ROSAT} All-Sky Survey (RASS). We are therefore able to provide much more general constraints on hot halo masses around $L*$ galaxies.

\section{Sample}

In choosing a sample of galaxies to stack, our highest priority was selecting very isolated galaxies. Galaxies in group or cluster environments are surrounded by hot, X-ray emitting gas in the form of the intragroup/intracluster medium, and it is far too difficult to distinguish a galactic hot halo from an intragroup medium using \emph{ROSAT}. Moreover, there could be complex interactions between this medium and the galactic hot halo which would complicate interpretation of the results.  We therefore chose the 2MASS Isolated Galaxy Catalog (2MIG, \citealt{Karachentseva2010}), which contains 3227 nearby galaxies selected using the following criteria: 1. detected in the 2MASS Extended Sources Catalog (XSC). 2. $K_S$ magnitude $< 12$. 3. angular diameter $a > 30"$. 4. the galaxy must be ``isolated'' from any other 2MASS XSC companions with $K_S < 14.5$.

We took the 3227 2MIG galaxies and narrowed the sample down to galaxies with measured radial velocities, and required the radial velocity to be at least $500$ km s$^{-1}$. This left  2496 galaxies in our total sample. Using their de Vaucouleurs classifications, we divided these galaxies into early-type (Elliptical and S0) and late-type (Sa-Sd, and irregulars), with 756 in the former and 1740 in the latter (including 7 irregulars). 

We also divided the galaxies at $L*$, which we took to be at $M*_{K_S} = -24.1$. This is the characteristic magnitude of the $K_S$ galaxy luminosity function (\citealt{Bell2003}; \citealt{Kochanek2001}), and roughly corresponds to the magnitude at which hot mode accretion is expected to dominate over cold mode accretion. To be precise, samples of nearby red and blue galaxies can have slightly different values for $M*$, but we ignored that small difference for this analysis. The luminous subsample has 1016 galaxies total, with 393 ellipticals and 623 spirals. The faint subsample has 1480 galaxies total, with 363 ellipticals and 1117 spirals. In Figure 1, we present histograms of our sample as functions of distance and absolute magnitude.

\begin{figure*}
\plottwo{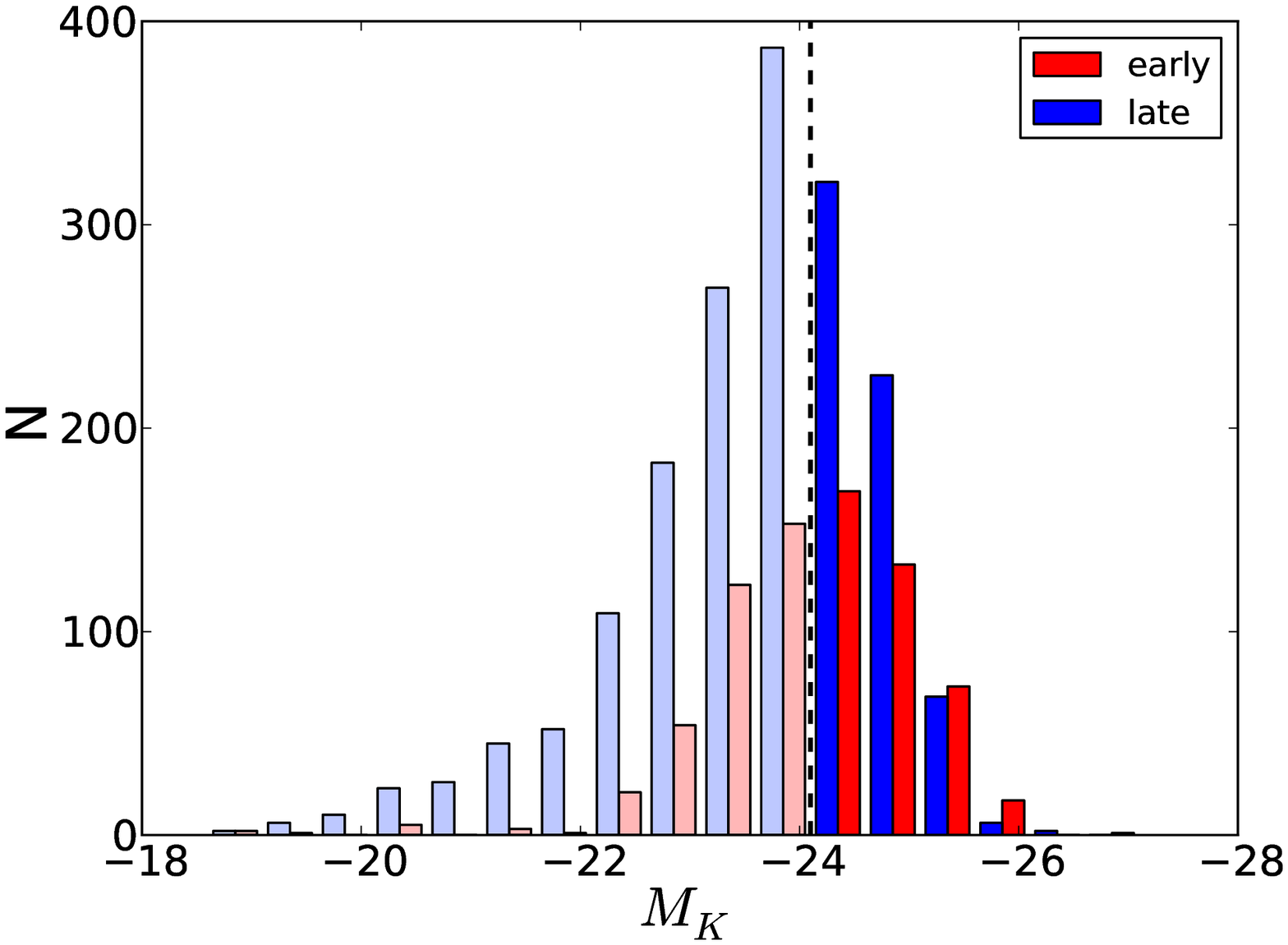}{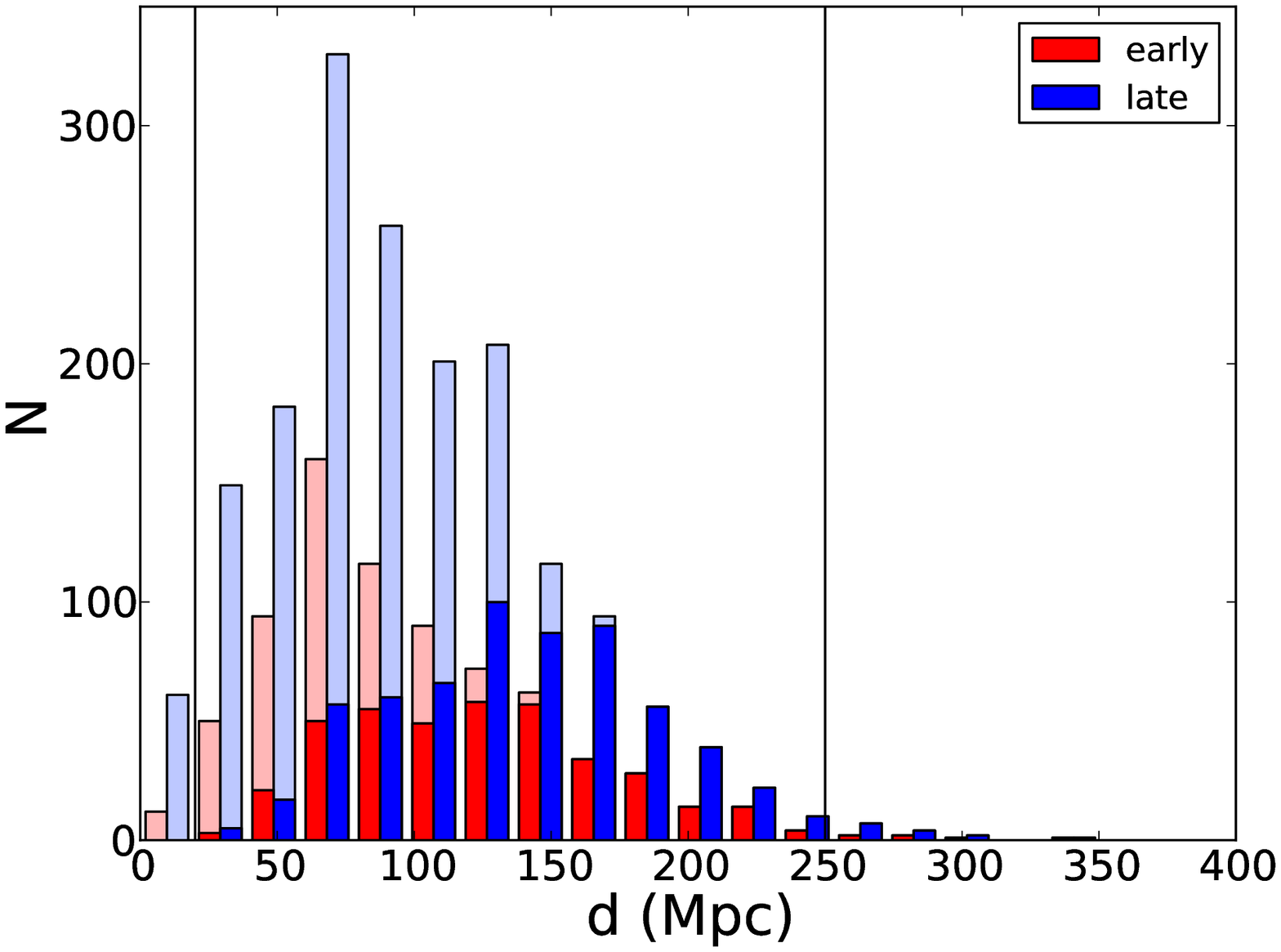}
\caption{Histograms of (left) $K_S$ absolute magnitudes and (right) distances for our sample of galaxies in the 2MASS  Isolated Galaxies catalog (2MIG, \citealt{Karachentseva2010}). We differentiate between the subsample of galaxies with $M_{K_S} < -24.1$ (indicated in the bolder colors) and the fainter subsample (indicated in the more muted colors). This characteristic magnitude is also denoted on the plot on the left with a black dashed line. We also place vertical lines at $d = 20$ Mpc and $d = 250$ Mpc; the 67 galaxies that fall outside this region are excluded from our stacking analysis. In total, we examine 2496 galaxies, divided into subsamples based on luminosity and morphological type, and after rejecting galaxies based on distance or point source contamination we eventually stack 2165 galaxies.\\ }
\end{figure*}

\section{Constructing Stacked Images}

The RASS offers several useful features for this sort of analysis.  The most important advantage is that the images have been carefully flattened \citep{Voges1999}, which reduces the effect of spatial variations in the background which could otherwise be confused with diffuse emission. The effect of the flat fielding is visible in the flatness of the background in the surface brightness profiles which we will present in this paper. Additionally, the RASS has nearly full sky coverage, so we have a large number of objects available for our sample, and can therefore measure down to extremely low surface brightness. The major disadvantage of using the RASS is its poor angular resolution (for which we compensate by constructing an empirical PSF, see section 3). 

To stack the galaxies in our sample, we used a modified version of the stacking code written by \citet{Dai2007} to stack RASS images of nearby galaxy clusters. We add all the photons in the 0.5-2.0 keV range within 500 projected kpc of each galaxy (each galaxy has a known redshift listed in  \citealt{Karachentseva2010}). Since each galaxy is at a different distance, each image therefore covers a different angular size on the sky, and we re-scale each image to the same angular size while preserving the photon count within 500 kpc.  

Since 500 kpc covers a larger angular size for nearby galaxies, this naturally leads to the nearby galaxies contributing more background photons per image than the more distant galaxies. Most of the galaxies in our sample lie around 100 Mpc, but the handful of images corresponding to much nearer galaxies can dominate the total stacked image. We therefore imposed a distance cut, excluding any galaxies with an angular diameter distance of less than 20 Mpc. This removes 67 galaxies from the sample. We also imposed a cut on the distant end ($d > 250$ Mpc) because the 20 galaxies at larger distances were contributing so few photons to the stack that we were adding noise. 

We also construct a stacked image of the RASS map of exposure time for each observation, again extending out to 500 projected kpc for each galaxy. We weight each individual exposure map by the quantity $($ 100 Mpc $/d)^2$ in order to account for the different angular sizes.  This exposure map is necessary because some galaxies lie near the edge of a RASS $6.4^{\circ} \times 6.4^{\circ}$ frame, and so we are missing some counts within 500 projected kpc on one side of these galaxies. This produces a sort of ``effective vignetting'', which we can model with knowledge of the exposure map. The total weighted exposure time is also useful in later analysis for converting the total measured counts into flux and luminosity units (section 7.1). 

We also cross-match each frame with the \emph{ROSAT} bright source catalog (BSC, \citealt{Voges1999}). We exclude any point sources in our frames that are listed in the BSC. After masking regions with known point sources, we randomly populate the masked regions with artificial photons matching the photon density of nearby regions. We also visually inspected each individual frame, and removed frames with extremely bright point sources that were not masked by our software. In total, there were 81 such frames. There were also 183 galaxies that were more than $190'$ from the center of the nearest RASS frame, and therefore rejected from the stacking algorithm (the frames are squares with sides of length $384'$). We create a stacked image of all remaining 2165 galaxies, as well as images of four subsamples. Two subsamples are divided by morphology, with 659 early-type galaxies and 1506 late-type galaxies. The other two subsamples are divided by luminosity, with a cutoff at $M_{K_S} = -24.1$, such that there are 911 luminous galaxies and 1254 faint galaxies. The resulting stacked images are shown in Figure 2. We binned the images into ``pixels'' of 5 projected kpc (10.3") on a side, and computed the average surface brightness in 1-pixel bins, displayed in Figure 3.

\begin{figure*}
\hspace{1.5 in}
\includegraphics[clip=true,width=6 in]{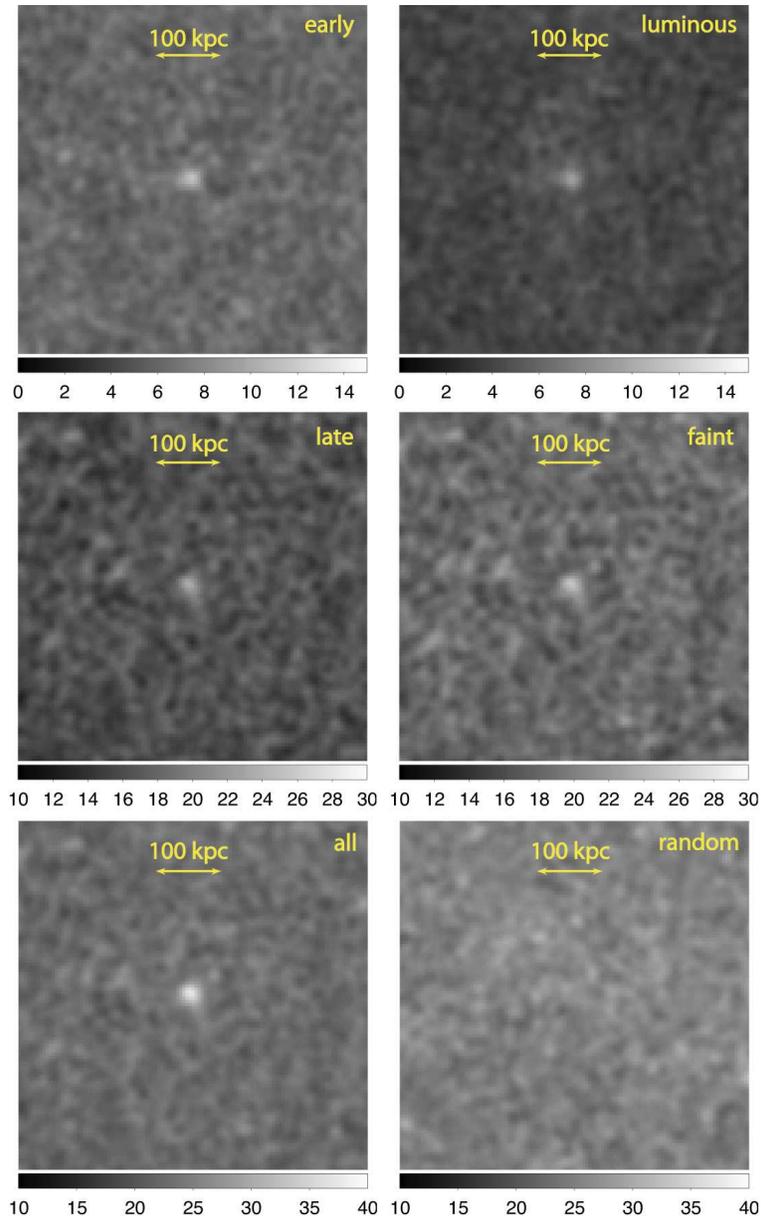}
\caption{ Stacked images of our samples of 2MIG galaxies.  Going clockwise from upper left, the samples are: early-type galaxies, luminous galaxies, faint galaxies, random positions on the sky (i.e., a null sample), all 2MIG galaxies, and late-type galaxies. All images have been smoothed with a 3-pixel Gaussian kernel. Note the different colorbars used in the three rows of images. Emission is clearly visible in the center of all the samples of 2MIG galaxies, and no emission is visible in the stack of random positions on the sky.\\}
\end{figure*} 

\begin{figure*}
\plotone{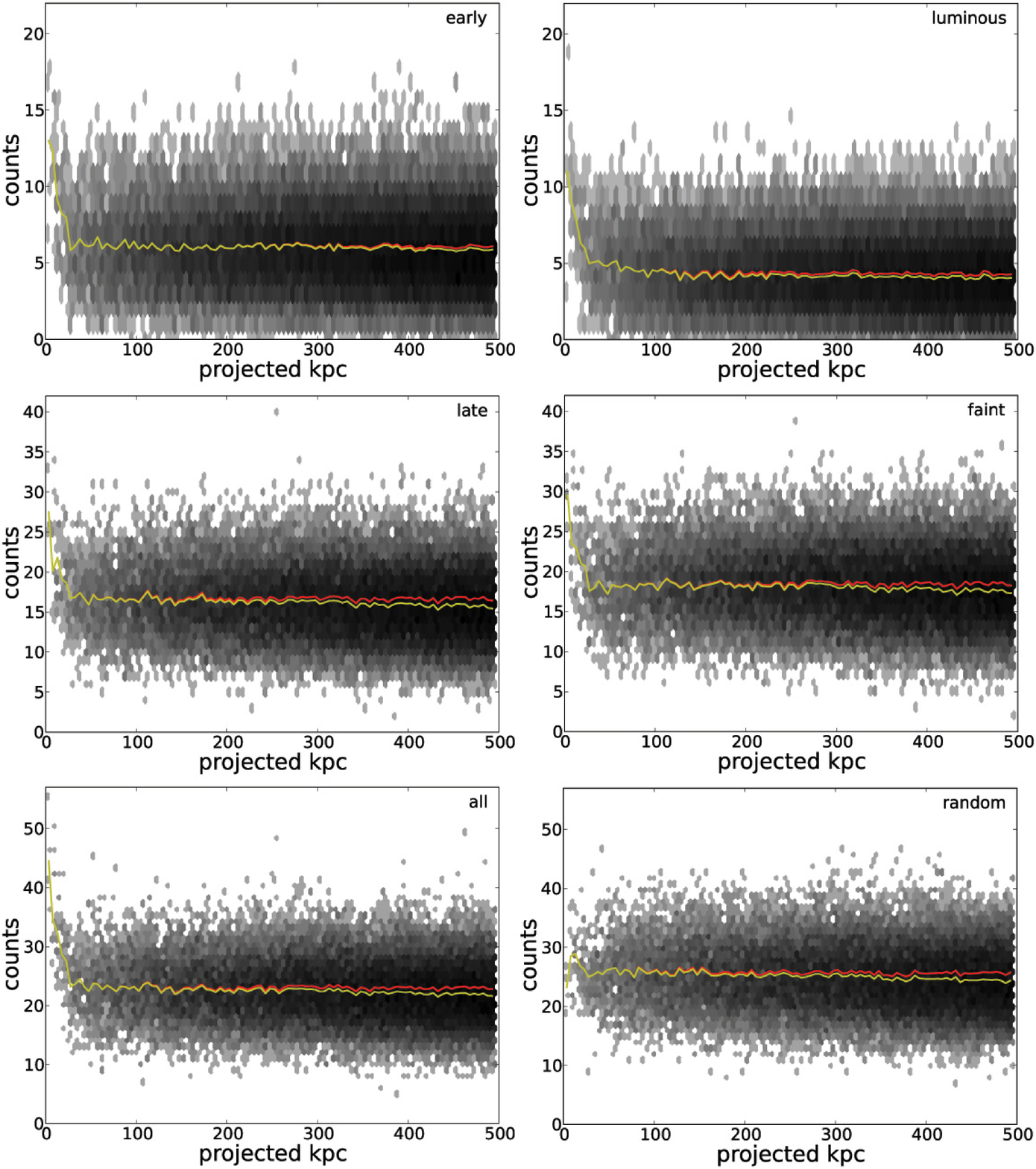}
\caption{ These plots contain the same information as the images in Figure 2, expressed as surface brightness profiles instead of images. The shading shows the number of counts per 5-kpc pixel, as a function of radius. The shading increases with radius because there are more pixels at large radii. The yellow line shows the mean surface brightness. The red line shows the mean surface brightness after dividing the mean by the average effective exposure at that radius. The difference between the red and yellow lines represents the effective vignetting described in section 3. \\}
\end{figure*}

These images generally show fuzzy emission in the central region, but we need to consider the point spread function for the \emph{ROSAT} PSPC to determine to what extent the emission is more diffuse than a point source. The off-axis PSPC PSF is fairly well-understood (with an analytic approximation available\footnote{\noindent http://heasarc.gsfc.nasa.gov/docs/journal/ROSAT\_off-axis\_psf4.html}), but our image is a composite of a variety of images at different off-axis angles. Moreover, the all-sky survey was conducted in scanning mode, so each source has been observed at a variety of (unknown) off-axis angles. This makes it impractical to construct an analytic composite PSF, so we chose instead to construct an empirical composite PSF. 

We generated a sample of point sources from the \emph{ROSAT} Faint Source Catalog (FSC) and the BSC designed to match the observed flux distribution of the galaxies in our sample. We assumed the galaxies in a given sample all have similar X-ray luminosities, and so the flux distribution is determined by the distribution of distances to the stacked galaxies. We created point source distributions for three sets of galaxies: the full sample, the luminous sample, and the faint sample. For each sample, we also examined the galaxies between 30 and 150 Mpc (a range which contains most of the galaxies in each case, but only produces a factor of 25 range in flux instead of 156 from the full 20-250 Mpc sample). For each galaxy in each sample, we picked the ten point sources in the FSC and/or the BSC with the fluxes that were closest to this galaxy's position in the distance-squared distribution. We assigned these point sources the redshift of the corresponding galaxy (in order to convert the stacked PSF into the same physical units we use for the real analysis), and stacked all the point sources. After rejecting a handful of outliers with other bright point sources in the frame, we have empirical PSFs  corresponding to the stacks of the full 2MIG sample, the faint galaxies, and the luminous galaxies, and containing 15440, 10564, and 4833 stacked point sources, respectively. In Figure 4, we present the radial surface brightness profiles for our fiducial PSFs. Note that our empirical PSF has a similar angular size to the PSF measured by \citet{Dai2010}.

This method is subject to a Malmquist bias, because within each subsample the more distant galaxies will be on average slightly more luminous than the nearer galaxies. This is a higher-order correction which we do not apply to our data because we do not know the intrinsic X-ray luminosities of our target galaxies (only their redshifts and K-band luminosities). Instead, we construct alternate empirical PSFs by stacking 25,000 point sources at random from the FSC, and assigning them redshifts of 2MIG galaxies at random. This is essentially an overcorrection for the Malmquist bias because we are effectively assuming here that every galaxy has the same X-ray flux and therefore we are over-weighting the more distant galaxies. Still, the resulting empirical PSFs are fairly similar to the fiducial PSFs; they have slightly more power in the 200-300 kpc range, and less power at larger radii, but the shapes of the cores are very similar. We ran the entire subsequent analysis with these alternate PSFs and the results were nearly identical. 

We also constructed another set of empirical PSFs by stacking 1946 K-type stars from the Michigan Spectral Catalogue (Houk et al. 1975). We used the same technique for stacking these images as for the FSC sources. The MSC is potentially less likely to contain extended sources, however, compared to the FSC, so it provides a valuable cross-check of our estimated PSF. On the other hand, the MSC sources are generally fainter than the FSC, and we have fewer than 10\% as many sources, so the PSF constructed from the MSC is not sampled nearly as fully. Still, these PSGs were very similar to the FSC PSFs above, further demonstrating the robustness of our determination of the PSF.

\begin{figure}
\epsscale{1.1}
\plotone{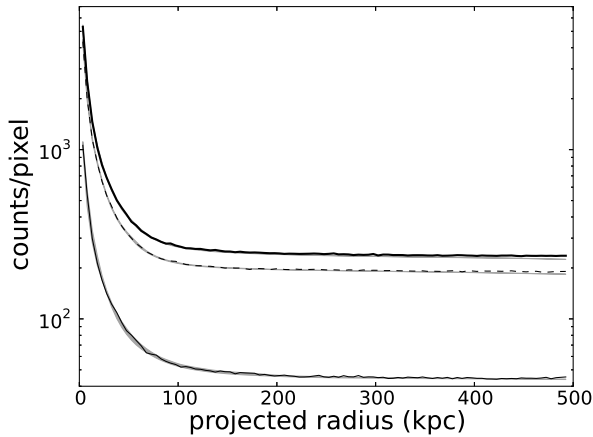}
\caption{Mean surface brightness profiles of the primary stacked PSFs, as described in section 3, without the ``effective vignetting'' correction. The thick solid line represents the surface brightness profile computed by assigning each point source a redshift at random from the full distribution of 2MIG galaxies. The thin solid line is the PSF that results if redshifts are assigned at random from only the luminous 2MIG galaxies, and the dashed line is the PSF that results if redshifts are assigned at random from only the faint 2MIG galaxies. Because the luminous galaxies are on average more distant than the full sample (and the faint galaxies on average are less distant), the aperture corresponding to 500 projected kpc includes more (fewer) photons, which affects the normalization of the PSF. The shaded regions around each surface brightness profile show the full range of MCMC fits to the data, using the model defined by eq. 4. \\}
\end{figure}

We also examined the effect of varying the distribution of redshifts assigned to the FSC sources. We will discuss this in section 5.4, but it turns out that using the wrong redshift distribution introduces an error of up to 30\% to the total flux. To avoid this error, we use the PSFs constructed with the appropriate redshift distributions in the subsequent analysis for these two samples.

\section{Analysis}

\subsection{Parameterizing the surface brightness profile}

The primary analysis in this paper is a Markov-Chain Monte Carlo fit to the various surface brightness profiles we constructed in the above section. An ``MCMC'' analysis efficiently samples likelihoods over parameter spaces of moderate dimensionality, yielding posterior probability distribution functions (pdfs) for each parameter of interest. It also easily allows us to marginalize over uninteresting parameters, and to compute pdfs for quantities derived from the surface brightness profiles. As it is a Bayesian technique, it requires specification of prior probability distributions for each parameter. In this analysis, we use the Python package pymc \citep{Pymc} to perform the MCMC fit, which uses a standard Metropolis-Hastings sampling algorithm. 

We begin by motivating the parameterization of the surface brightness profile. As Figure 3 shows, at large projected radii from the stacked galaxies (if we account for the ``effective vignetting'' discussed in section 3), the profile looks flat; we therefore assume the background is uniform across the detector, and define the parameter $A_{c}$ as the constant background in units of counts/pixel. In fact, the remarkable flatness of the background is one of the primary advantages of using the RASS for this analysis (in particular, note the flatness of the random profile in Figure 3). We assume that the extended emission from the galaxies can be parameterized with a $\beta$-model, $b(r)$, defined as 

\begin{equation}b(r) = \left[ 1 +\left(\frac{r}{r_0}\right)^2 \right]^{0.5-3\beta}\end{equation}

In this model, $r_0$ is the core radius, $\beta$ is the slope, and we pull the normalization $A_{\beta}$ out in front of the convolution. There are many advantages of using a $\beta$-model for the hot gas component. If we assume the gas is isothermal and has a constant metallicity (which we do assume for this analysis), then its density profile is given by

\begin{equation}n(r) = n_0 \left[1 + \left(\frac{r}{r_0}\right)^2\right]^{-1.5\beta}\end{equation}

which we can integrate to find the hot gas mass within a given radius. Empirically, these models fit fairly well to the hot gas around elliptical galaxies \citep{Forman1985} as well as the hot gas in galaxy groups and clusters \citep{Sarazin1986}, and we used them to fit the surface brightness profile around the isolated giant spiral galaxies NGC 1961 \citep{Anderson2011} and UGC 12501 \citep{Dai2011} as well. 

We also include an additional component intended to mimic the signal from the center of the galaxy and any active nucleus it may contain. We parametrize this with a $\delta$-function, $\delta(r)$, which is approximated as $A_{\delta}$ in the first pixel and 0 in the other pixels. Any emission which is restricted to the nucleus, the bulge, or in the inner 5 kpc of the stellar component is intended to be captured in this component. Then we convolve $b(r)$ and $\delta(r)$ with a PSF (using the symbol $*$ to denote a discrete convolution) to produce the full signal from the source. 

Finally, as mentioned above (and as can be seen in Figure 3), the surface brightness profile of the data declines slightly at large radius. This is because some of the images we stack do not extend to 500 projected kpc on each side of the galaxy, meaning that the final image is missing some photons at large radii. We account for this by measuring the surface brightness profile of the weighted, stacked exposure map for each sample, and normalizing by the central (maximum) exposure time to produce an effective exposure profile for each sample, $f_{\text{vig}}(r)$. At 500 projected kpc, $f_{\text{vig}}$ is still over 90\%, but this 10\% correction can still be important for measuring the background accurately. The final model is therefore: 

\begin{eqnarray}S(r) &=&   A_{\delta} \times  \left[\delta(r) * \text{psf}(r)\right] \times f_{\text{vig}}(r)  \nonumber\\
 &+&   A_{\beta} \times \left[b(r) * \text{psf}(r)\right] \times f_{\text{vig}}(r) \\
 &+&    A_{c}  \times f_{\text{vig}}(r) \nonumber \end{eqnarray}

\subsection{Computing the PSF}

To estimate PSFs from the surface brightness profiles in Figure 4, we need to subtract the uniform background component. We estimate the level of this background by parameterizing the PSF analytically and running an MCMC fit to the surface brightness profile to compute the posterior probability distribution for the level of the background. Motivated by the analytic expression for the off-axis PSF in targeted PSPC observations: \citep{Hasinger1993}

\begin{eqnarray}\frac{\text{psf}(r)}{f_{\text{vig}}(r)} &=& A_g \times \text{exp}\left(\frac{-r^2}{2 \sigma^2}\right) \nonumber\\
&+&  (r > r_c)\times A_p \left(\frac{r}{r_0}\right)^{-\gamma} \\
&+&  (r \le r_c) \times \frac{A_L}{1+(2r/r_L)^2} \nonumber\\
&+& A_c\nonumber \end{eqnarray}

So the PSF is defined by the sum of a Gaussian component, a Lorentzian component that becomes a powerlaw at large radii ($A_L$ is fixed so that the Lorentzian and the powerlaw have the same value at $r = r_c$), and a uniform background component. We use uniform priors on all parameters except $A_c$, which uses Gaussian priors centered around the mean value of the surface brightness profile at large radii (and standard deviation of 0.1). The powerlaw slope $\gamma$ is also required to be less than -2, in order to guarantee that the PSF converges to zero at infinitely large radii. The variable $r_0$ is just a normalization constant; we take $r_0 = 100$ pixels $= 500$ projected kpc. Note that we have moved $f_{\text{vig}}(r)$ to the left just to make the equation easier to read; $f_{\text{vig}}(r)$ is always applied to the model, not to the data, for the MCMC analysis. 

We multiply both the data and the model by the number of pixels in each 1-pixel radial bin, since we are actually measuring the total number of photons and not the average per radial bin. We then compute uncertainties in each bin according to Poisson statistics. We ran $3\times10^6$ iterations, discarded the first $5\times10^4$, and selected only every 1000th remaining iteration in order to ensure each element is independent of each subsequent element.  The results of the MCMC fitting are shown above, graphically in Figure 4, and the probability distribution functions (pdfs) of the fit parameters are listed in Table 1. 

\begin{deluxetable*}{cccccccc}  
\tablecolumns{8}
\tablecaption{PSF fitting parameters}
\tablehead{   
  \colhead{PSF name} &
  \colhead{$A_p$} &
  \colhead{$A_g$} &
  \colhead{$A_c$} &
  \colhead{$\gamma$} &
  \colhead{$\sigma$} &
  \colhead{$r_c$} &
  \colhead{$r_L$} 
}
\startdata
all & $0.46\pm0.03$ &$4366_{-73}^{+78}$& $239.8\pm0.1$ & $2.57\pm0.03$ & $0.80\pm0.01$ & $11.2\pm0.2$ & $5.5\pm0.1$\\
faint & $0.30\pm0.02$ & $3658_{-74}^{+76}$ & $193.7\pm0.1$ & $2.63\pm0.04$ & $0.76\pm0.01$ & $10.7\pm0.3$ & $5.2\pm0.1$\\
luminous  & $0.20_{-0.03}^{+0.02}$ & $877_{-27}^{+31}$ & $44.8\pm0.1$ & $2.32_{-0.06}^{+0.08}$ & $0.93\pm0.02$ & $12.3_{-0.8}^{+1.5}$ & $6.8\pm0.2$
\enddata
\tablecomments{Median values and 68\% confidence regions for the MCMC fits to the three psfs examined in section 4.2. Note that all of these parameters are computed in units of 5-kpc pixels, so to convert $r_c$ into kpc, for example, multiply it by 5.}
\end{deluxetable*}

Because the stacked FSC and MSC data (the latter in particular) are so noisy, we use the best-fit profiles from the MCMC analysis to generate the PSFs. We divide this best-fit profile by  $f_{\text{vig}}(r)$ to correct for the artificial vignetting, and we subtract the median value of $A_c$ to yield an empirical PSF. The mean value of the PSF surface brightness at large radii turns out to be slightly above $A_c$. The discrepancy is about 4\%, suggesting that this fraction of the signal at large radii comes from the wings of the PSF and not from the uniform background. 

In Figure 5a, we show the resulting PSFs, after subtracting off the background component and normalizing the result so the integrated sum is unity. We also show average power at each radius, multiplying the PSF at each radius by the number of pixels it subtends, in Figure 5b (the discrete pixel size causes the apparent noise). The two sets of point spread functions, although computed from different objects and using very different sample sizes, have similar shapes. The cores seem weaker in the PSFs constructed by assigning redshifts from the sample of luminous galaxies, compared to the other PSFs (Fig. 5a, inset), and the PSFs are less well constrained at large radii where there is very little signal, but there are not significant systematic differences otherwise. 

Considering how much smoother the underlying data are, we adopt the FSC PSFs for subsequent analysis. The fiducial FSC PSF is used to analyze the full sample and the early-type and late-type subsamples. The FSC PSF constructed with the redshift distributions of the faint and the luminous galaxies are used to analyze these two subsamples, respectively.

\begin{figure*}
\plottwo{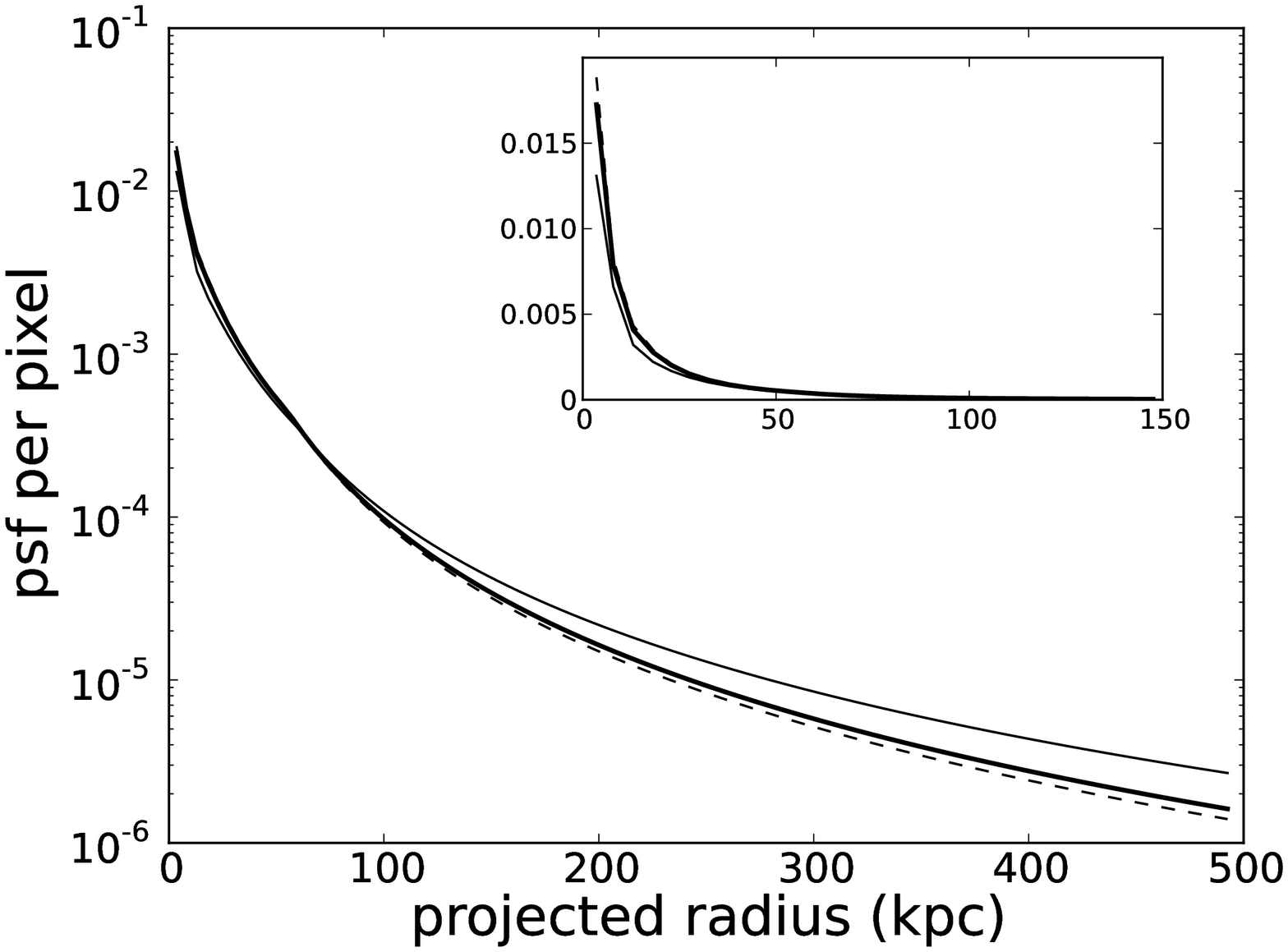}{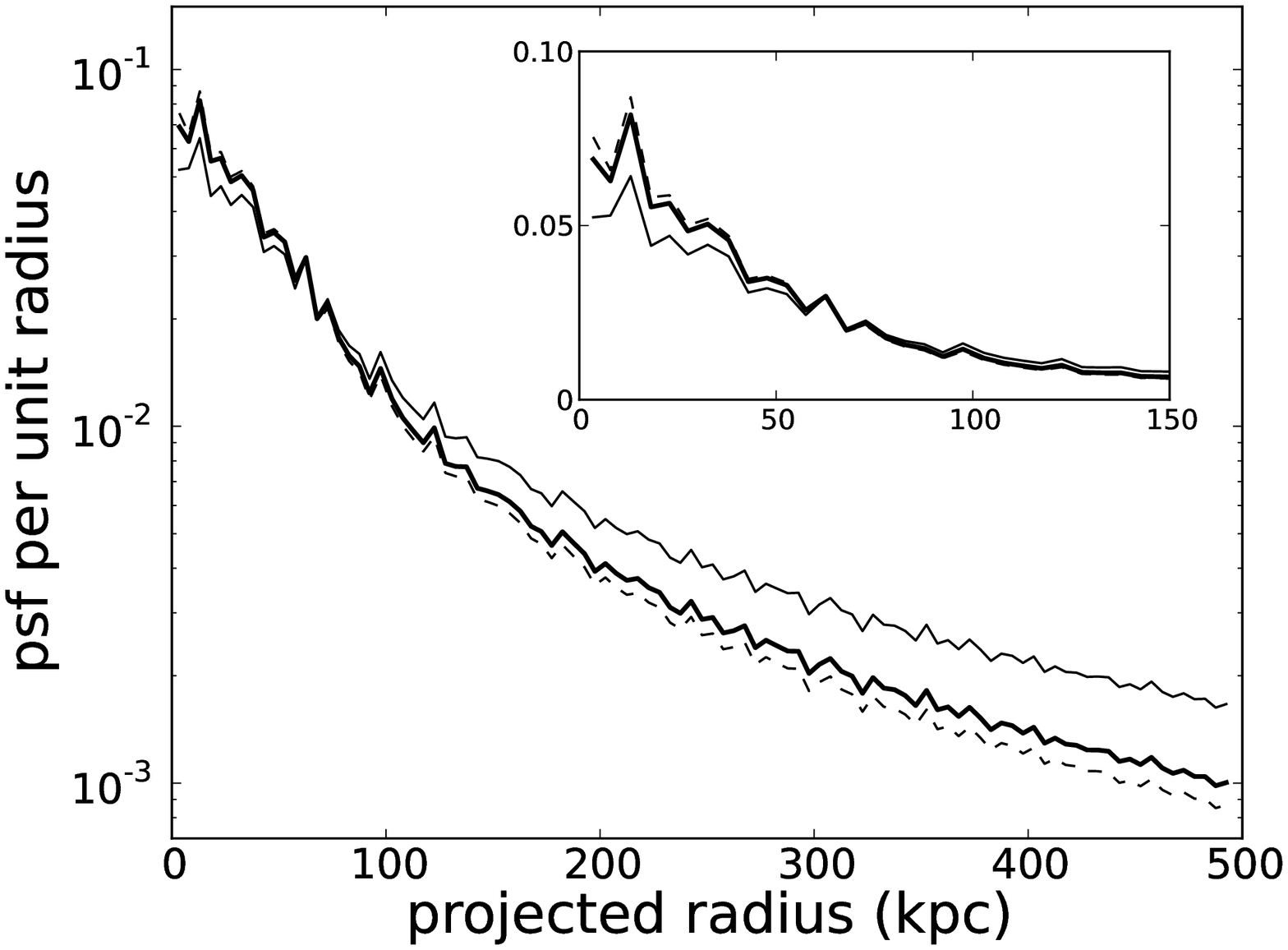}
\caption{ Profiles showing the normalized strength of the PSF as a function of projected radius. In (a), we show the PSF value per 5-kpc pixel, normalized so the total across the image adds to unity. In (b) we show the PSF value per 5-kpc radial bin (i.e., the PSF weighted by the azimuthal area it subtends), again normalized so the total out to 500 kpc adds to unity. The inset in each plot shows the same data on a linear scale. In each plot, the thick solid line corresponds to the PSF constructed from the redshift distribution of all 2MIG galaxies, the thin solid line is the PSF constructed from redshifts of only the luminous galaxies, and the dashed line is the PSF constructed from redshifts of only the faint galaxies. \\ }
\end{figure*}

\section{Simulations}

Now that we have a functional form for the PSF, we are able to begin analyzing the stacked 2MIG images. However, to verify the reliability of our stacking technique and to aid in the interpretation of the results, we first apply our analysis to simulated stacked images.  We constructed the simulated images based on the model in equation 3. For each image, we specified all of the free parameters in equation 3 (i.e. $A_{\delta}$, $A_{\beta}$, $A_c$, $r_0$, $\beta$, the functional form of the PSF, and the exposure map from the stack of real 2MIG galaxies corresponding to the simulated image). For each of the three components of the model (the point source, the extended component, and the uniform background - note that the first two components are each convolved with the specified PSF as well), we computed the expected number of counts in each 5 kpc $\times$ 5 kpc ``pixel'', and then placed in that pixel a number of counts equal to a random Poisson deviate with the given expected value. We added the three components together and multiplied by the normalized exposure map to produce a full simulated image with the same ``effective vignetting'' as the real image.

In the rest of this section, we present and discuss a number of  simulations. Note that instead of $A_{\delta}$ and $A_{\beta}$, we found it more useful in analyzing the simulations to specify $N_{\delta}$ and $N_{\beta}$ - the total number of counts in the point source and extended components, respectively. Recovering $N_{\delta}$ and $N_{\beta}$ from a simulated image is our first priority in this analysis. We define the total number of counts within a circle of radius 500 projected kpc, or 100 pixels. This means that, in theory, we could also estimate the number of counts $N_c$, which should be within integrated pixel-by-pixel Poisson errors of $10^4 \pi A_c$. However, in every simulation we run, $N_c$ is much larger than either $N_{\delta}$ or $N_{\beta}$, and we find it conceptually easier to work with $A_c$. Table 2 lists the basic properties of our simulated images. The only differences between the various sets of simulations are the quantities listed in Table 2.

\begin{deluxetable}{ccccc}  
\tablecolumns{5}
\tablecaption{Properties of simulations}
\tablehead{   
  \colhead{name} &
  \colhead{$N_{\delta} + N_{\beta}$} &
  \colhead{$A_c$} &
  \colhead{PSF} &
  \colhead{exposure map}
}
\startdata
All galaxies & 1600 & 22.6 & all & all\\
Late-type galaxies& 800 & 16.6 & all & late\\
Early-type galaxies& 800 & 6.0 & all & early\\
Faint galaxies & 600 & 18.4 & faint & faint\\
Luminous galaxies &1000 & 4.2 & luminous & luminous\\
Extra counts& 10000 & 22.6 & all & all
\enddata
\tablecomments{Basic properties of the sets of simulated images we examine. The simulated images are designed to mimic the properties of the real stacked images  of each subsample of 2MIG galaxies. The $N_{\delta} + N_{\beta}$ column shows the number of source counts spread across the image, and the numbers are chosen to approximate the estimated number of source counts in the corresponding real image. The $A_c$ column shows the average number of background counts per pixel, and again these numbers are chosen to approximate the estimated background in the corresponding real images. The PSF column shows the redshift distribution of 2MIG galaxies used to produce the PSF for the image; these are the same PSFs used to simulate the real images. The exposure map column shows which real stacked exposure map we use to simulate the ``effective vignetting'' in the simulated images. Finally, note that ``Extra Counts'' simulations have more photons than any real image and are designed to reduce the Poisson uncertainties and therefore help diagnose which errors in the simulations stem from photon noise and which errors are systematic.}
\end{deluxetable}

We analyzed these simulated images using the same pipeline as the real stacked images. This means for each simulation we specify the appropriate FSC PSF and the stacked exposure map corresponding to the image (see Table 2). The choices of priors for the model parameters are described below, in section 7. The only prior that uses information about the image is $A_c$, to which we assign a Gaussian prior centered at the the average number of counts per pixel at large radii, after attempting to correct for the artificial vignetting. We tested the choice of this prior, and it converges to the same value if we move the mean upwards or downwards slightly, or if we widen the width of the Gaussian. After running the MCMC analysis, we examined the resulting posterior probability distributions for $\beta$ and $r_0$, $A_{\beta}$, and $A_{\delta}$. From the first three of these distributions, we computed the posterior probability distribution for $N_{\beta}$, and from the fourth distribution we computed $N_{\delta}$.

\subsection{Recovering total counts and fit parameters}

We first examine the simplest case, where we set either $N_{\beta}$ or $N_{\delta}$ to zero. In Figure 6 we compare the pdfs (we show the medians and 68\% confidence intervals) for $N \equiv N_{\beta} + N_{\delta}$ to the true value for the simulations of each galaxy sample. We integrate the models over the entire image for this plot (out to 500 kpc); as we will show, it is easier to recover the emission just within 50 kpc or so, where it is significantly above the background. Still, across the whole image we are able to recover $N_{\beta} + N_{\delta}$ adequately for entirely point source emission and for $\beta > 0.5$ (i.e. extended emission with a fairly steeply declining slope). For flatter extended emission, the uncertainty grows, but in general the MCMC analysis fails to recover all of the emission, and the discrepancy increases as the slope becomes flatter. What seems to be happening is that the flatter profiles are becoming indistinguishable from Poisson noise at larger radii, and so the MCMC analysis is attributing the missing counts to a slightly higher background instead of to the extended component. 
 
\begin{figure}
\plotone{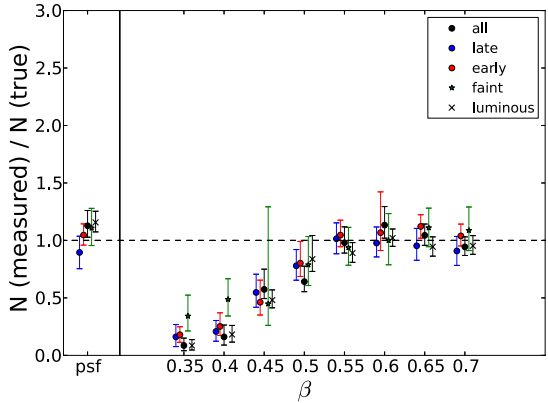}
\caption{ Probability distribution functions (pdfs) of the ratio of the measured value of $N \equiv N_{\beta} + N_{\delta}$ to the true value of $N$ for simulations of each galaxy sample. We measure $N$ using the MCMC analysis to fit the simulated image, and we integrate the resulting models out to the edge of the image (500 kpc). The true values of $N$ are listed in Table 2. 68\% confidence intervals around the medians of the pdfs for $N$ are shown as well. The points labeled 'psf' are simulations where all the source counts are in the point source component. All the other points are simulations where all the source counts are in the extended component, the slope $\beta$ of the extended component is indicated for each set of simulations. The points correspond to various simulations, as indicated in the legend. The points have been scattered by up to $\beta = 0.01$ for readability; each set of simulations actually uses the same values of $\beta$ at the indicated values of $\beta$. For discussion of this plot, see the text (section 5.1).\\ }
\end{figure} 

To help verify this, we checked if the MCMC analysis was able to recover the value of $\beta$ correctly (Figure 7). For the simulation with extra counts, $\beta$ is recovered approximately, but the other stacks almost always recover a value of $\beta$ around $0.5-0.6$, regardless of the true value. Unfortunately, this means we cannot use the recovered value of $\beta$ as a reliable guide to the true value of $\beta$. This also suggests that measurements which rely on extrapolating with $\beta$ out to large radii are unreliable if we suspect $\beta \lapprox 0.5$.

\begin{figure}
\plotone{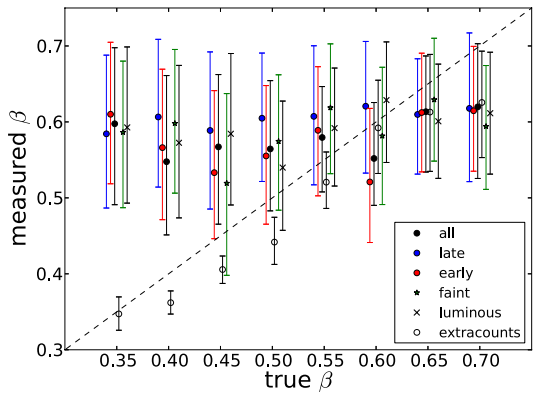}
\caption{ Probability distribution functions (pdfs) for the measured value of the slope $\beta$ of the extended component of the simulated images, compared to the true value of $\beta$ for those images. The 68\% confidence intervals are shown for each simulation as well. The dashed line denotes equality, i.e. where the measured value equals the true value. Points are colored/shaped as indicated in the legend. The points have been scattered horizontally by up to $\beta = 0.01$ for readability; each set of simulations actually uses the same values of $\beta$ at the indicated values of $\beta$ on the horizontal axis. This plot shows that, while with more counts it would be possible to recover $\beta$ approximately, we are not able to recover $\beta$ in simulations corresponding to our actual data. \\}
\end{figure} 

\subsection{Recovering counts within 50 kpc}

While it is difficult to distinguish a flattened profile from the background at very large radii, this emission is stronger at smaller radii, and therefore should be easier to recover. In Figure 8 we show the pdf for $N$ if we only integrate the fit parameters out to 50 kpc instead of out to 500 kpc. We compare the recovered counts within this radius to the expected number of counts within this radius, which can be computed easily since we know $\beta$ for our simulations (Table 2). Note that the number of counts within 50 kpc varies significantly with $\beta$, ranging from 11\% of the counts within 500 kpc for $\beta = 0.35$ to 94\% of the total counts for $\beta = 0.70$.  As Figure 8 shows, we are able to recover the number of counts within 50 kpc for most cases. The uncertainties appear much larger for the flatter profiles, but this is mostly because the total number of counts within 50 kpc is so much smaller. 

Additionally, there appears to be a slight trend of overestimating $N$ within 50 kpc, for $\beta \approx 0.5$, in the images with all the emission in an extended component. This happens because some extended emission is mistakenly attributed to the point source component, which increases the inferred number of counts originating from radii within 50 kpc.

\begin{figure}
\plotone{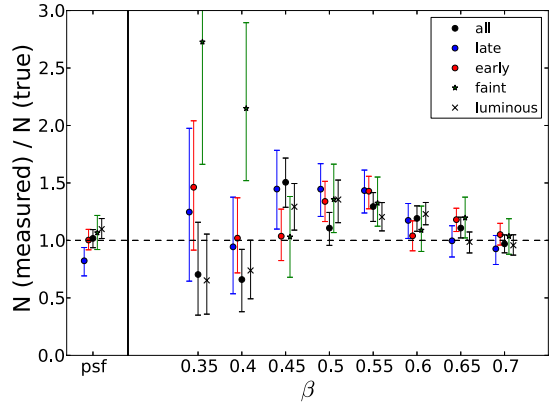}
\caption{ Probability distribution functions (pdfs) of the ratio of the measured value of $N \equiv N_{\beta} + N_{\delta}$ to the true value of $N$ for simulations of each galaxy sample. We measure $N$  using the MCMC analysis to fit the simulated image, and we integrate the resulting models out to a radius of 50 kpc. The true values of $N$ range from 11\% of the total listed in Table 2 to 94\% of the total listed in Table 2, depending on the slope of the profile. 68\% confidence intervals of the pdfs for $N$ are shown as well. The points labeled 'psf' are simulations where all the source counts are in the point source component. All the other points are simulations where all the source counts are in the extended component, the slope $\beta$ of the extended component is indicated for each set of simulations. The points correspond to various types of simulations, as indicated in the legend. The points have been scattered by up to $\beta = 0.01$ for readability; each set of simulations actually uses the same values of $\beta$ at the indicated values of $\beta$. We are able to recover the number of counts within 50 kpc for most cases. The uncertainties appear much larger for the flatter profiles, but this is primarily because the total number of counts within 50 kpc is so much smaller.\\ }
\end{figure}

\subsection{Identifying extended emission}
We now examine how well the MCMC analysis can distinguish point source emission from extended emission. In other words, we have examined how well we can recover $N \equiv N_{\beta} + N_{\delta}$, and now we examine how well we can recover $N_{\beta}$ and $N_{\delta}$ individually. To do this, we focus on the simulations of the stack of all 2MIG galaxies. In Figure 9 we show the recovered counts within 50 and 500 kpc, as well as the pdfs for $N_{\beta}$ and $N_{\delta}$. There are two separate but related failure modes. First, there is a constant -- but fairly small -- level of confusion is each simulation; the emission should be entirely in one component, but some emission is always attributed to the other component as well. Because all the point source emission is by definition within 5 kpc, the point source comprises a larger fraction of the true signal within 50 kpc compared to 500 kpc, so this failure mode hurts our ability to measure the correct fraction of extended emission within 50 kpc, and it has less of an effect at 500 kpc. Second, at $\beta \gapprox 0.6$, the MCMC analysis is unable to distinguish extended emission from point source emission, so it mistakenly attributes most of the emission to the point source. This is because the PSF is wider than the profile at $\beta \gapprox 0.5$, so the convolution of the two looks very similar to the PSF.  This failure mode is still quite significant at 500 kpc, because the steeper profiles have very little emission at larger radii. 

\begin{figure*}
\plottwo{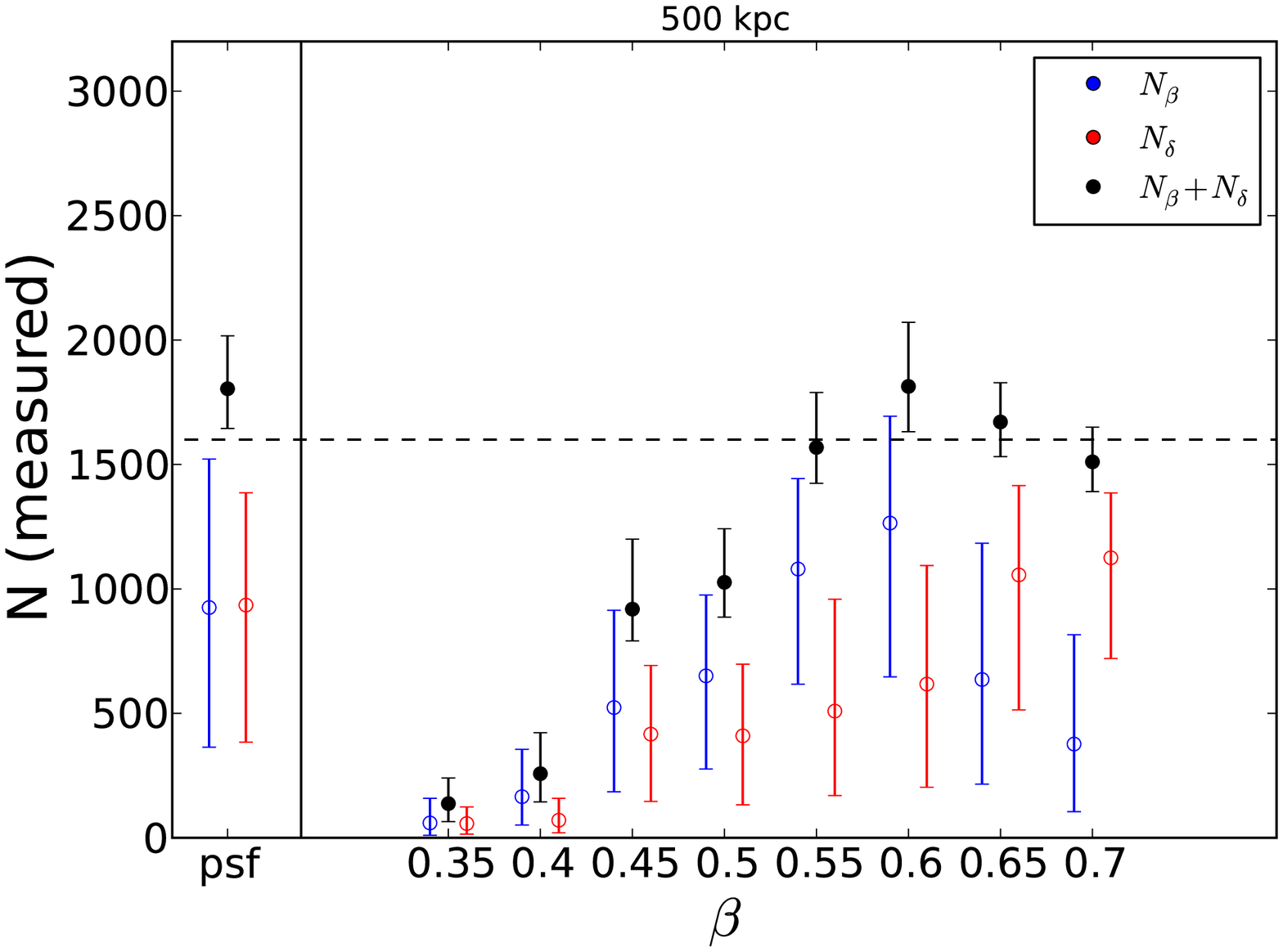}{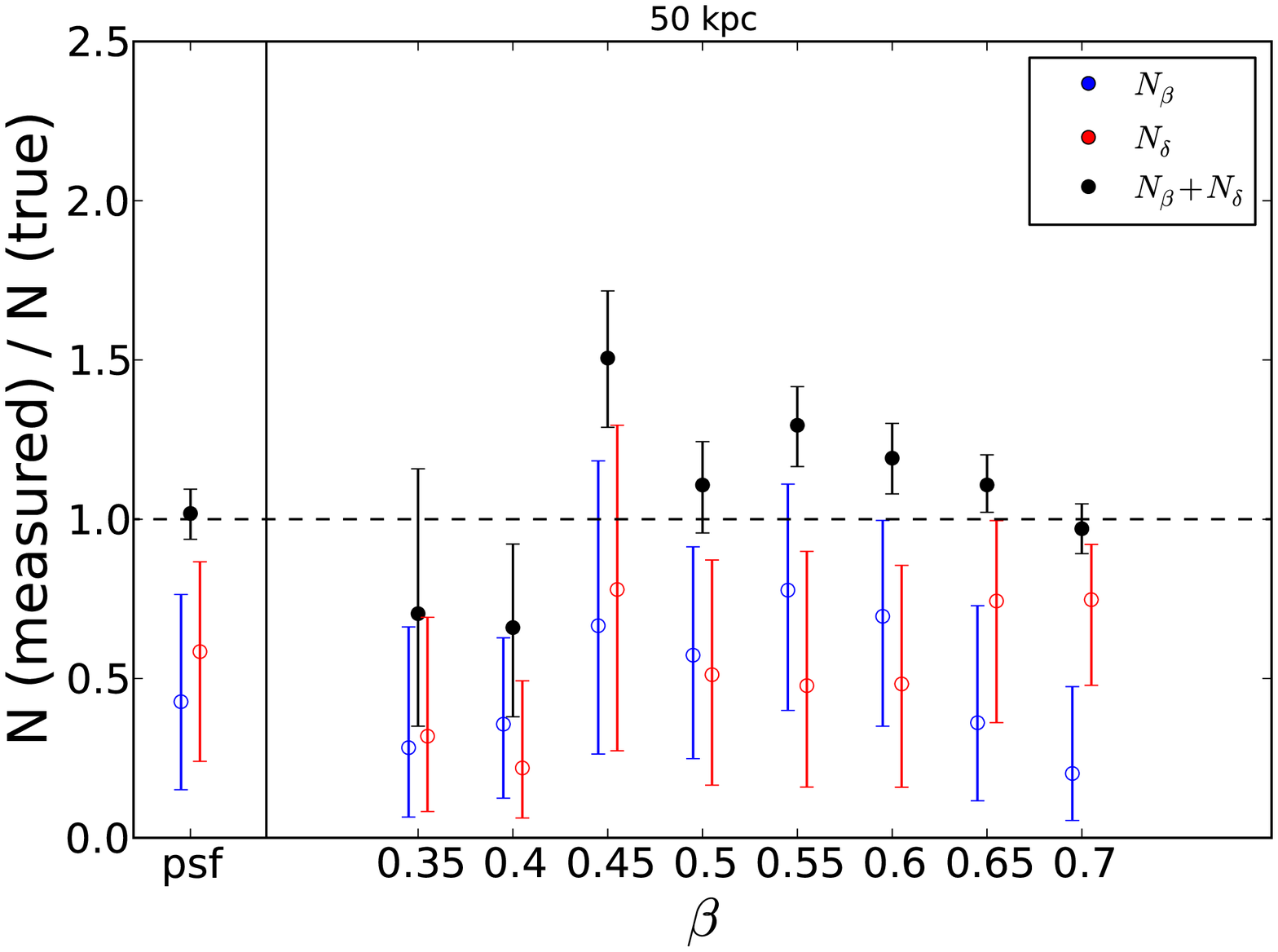}
\caption{ Probability distribution functions (pdfs) for (a) the values of $N_{\beta}$ and $N_{\delta}$ integrated out to 500 kpc and (b) the ratios of the values of $N_{\beta}$ and $N_{\delta}$ integrated out to 50 kpc compared to the true values (dashed lines). 68\% confidence intervals on each measured value are shown as well. The blue points denote the pdf for the extended component $N_{\beta}$, the red points the pdf for the point source component $N_{\delta}$, and the black points the pdf for the total source counts $N = N_{\beta} + N_{\delta}$. These pdfs are all measured for the ``all'' simulations. In the leftmost simulation, the true source is entirely composed of point source emission, and the other simulations have sources entirely composed of extended emission. Unfortunately, the MCMC analysis does not distinguish between the two types of emission reliably: for steeper profiles ($\beta \gapprox 0.5$) the extended emission is mistaken for point source emission, and all simulated images have some emission mistakenly attributed to the wrong component as well. Later in the paper, we combine this spatial analysis with a likelihood ratio test and a spectral analysis to improve our constraints on the extended component.  }
\end{figure*} 

The failure modes identified above - failure to recover counts at large radii if $\beta$ is too low, inability to distinguish point source emission from extended emission if $\beta$ is too high, and some confusion between point source and extended emission at all values of $\beta$ - could be due to insufficient photons or to systematic failures of the method. We attribute the first failure mode to insufficient photons (and note that this conclusion is supported by the fact that the ``all'' simulation, which has over twice as many source photons as the other simulations, better recovers the number of photons in the cases with low $\beta$), and the second failure mode to a fundamental systematic uncertainty: the poor angular resolution afforded by the PSF. To help support this claim, we now examine the ``extra counts'' simulations. These images have the same background and exposure map as the stack of all 2MIG galaxies, but we have 10000 source counts instead of 1600. The results are shown below, in Figure 10.

\begin{figure*}
\plottwo{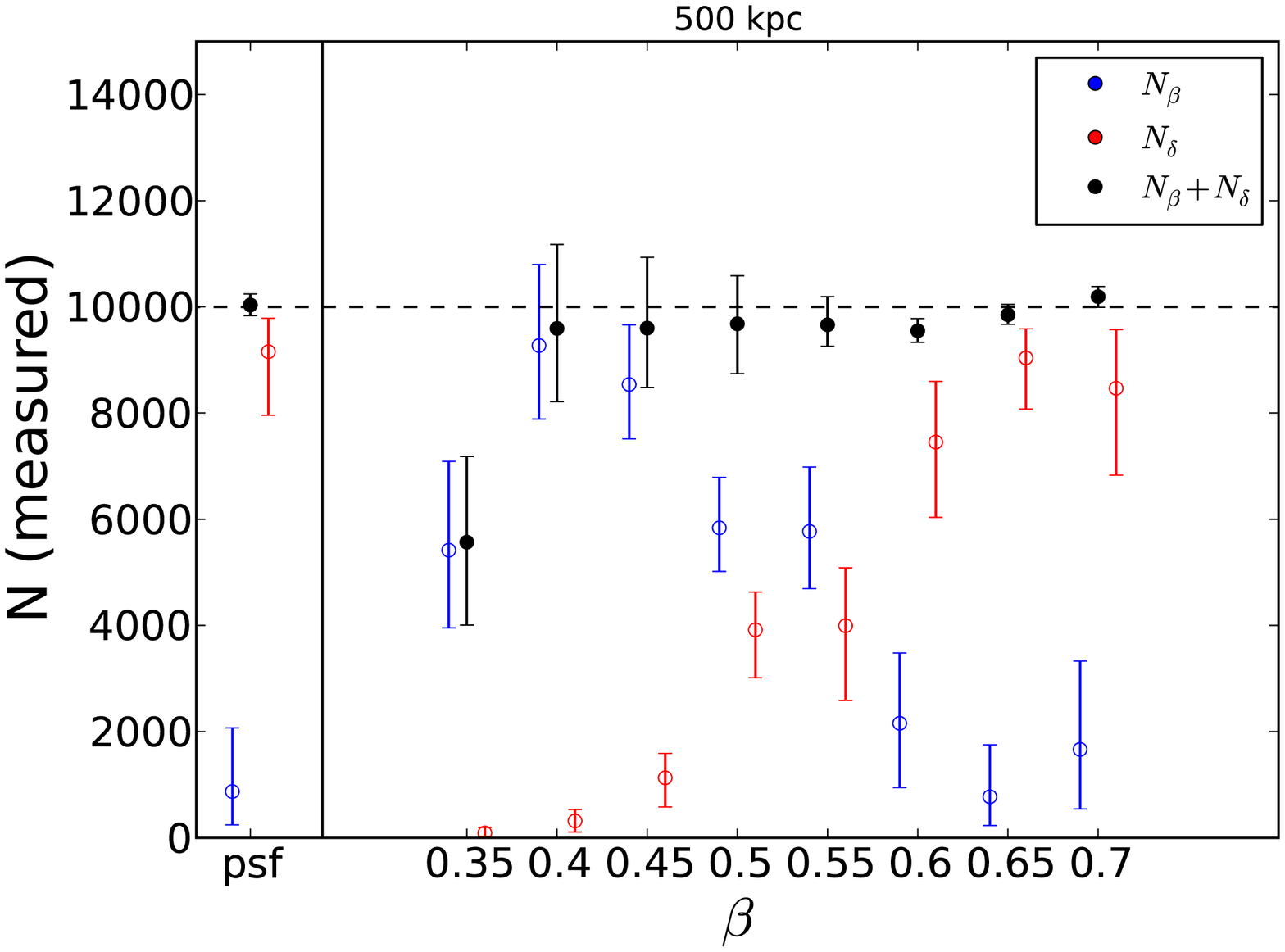}{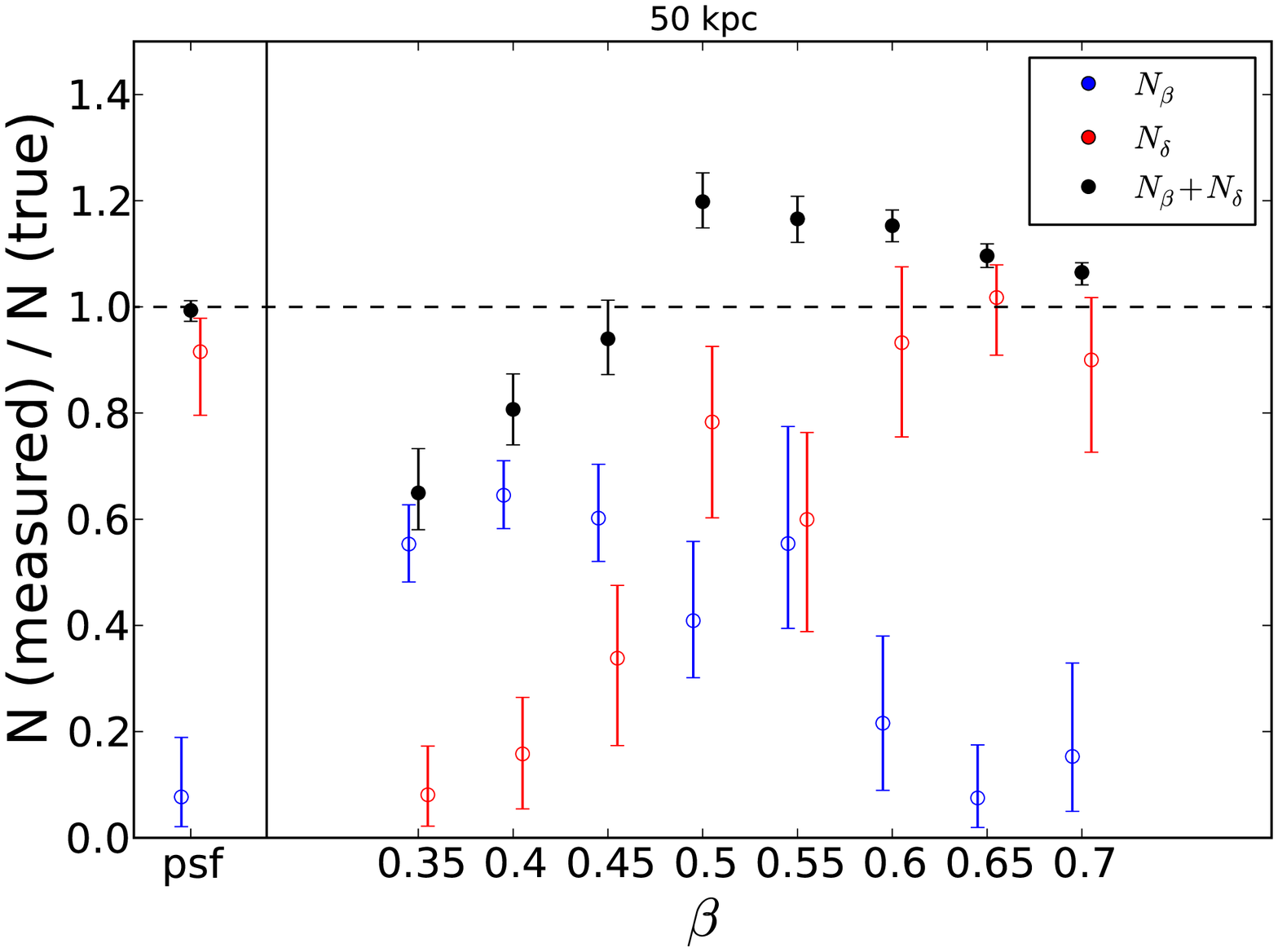}
\caption{ Same as Figure 9, but for the ``extra counts'' simulations. The MCMC analysis is still not able to distinguish between the two types of emission reliably for $\beta \gapprox 0.5$, but the amount of misclassified emission is much smaller in the other profiles, compared to the ``all'' simulations (Figure 9). Later in the paper, we combine this spatial analysis with a likelihood ratio test and a spectral hardness ratio analysis to improve our constraints on the extended component. }
\end{figure*} 

This Figure shows that the total number of counts ($N_{\beta} + N_{\delta}$) can be recovered to within 10\% for $\beta \ge 0.4$ within 500 kpc. So failures to recover this parameter are due to insufficient photons to distinguish the surface brightness profile from Poisson variations in the background. However, these simulations still confuse extended emission and point source emission for $\beta \gapprox 0.5$, so this seems to be a systematic failure of the analysis: the PSF is broad enough that the steeply declining extended profiles do not look very different from a point source. This also helps to explain why $N$ is only recovered to 30\% accuracy within 50 kpc: the misattribution of extended emission to a point source component increases the inferred number of counts at relatively small radii, although the total number of counts across the full 500 kpc image is still recovered correctly. The same issue applies in Figure 9b, although the statistical uncertainties are larger so this systematic effect is less important. 

Finally, the amount of confusion between point source and extended emission at lower values of $\beta$ and in the case of pure point source emission is much lower than it was in Figure 9. This implies that this confusion is due to insufficient photons - the MCMC analysis does work in the cases where the extended emission looks different from a point source. Thus, these failure modes should not affect the stacking analysis of high S/N data such as galaxy clusters (e.g. \citealt{Dai2010}).

\subsection{Effects of changing the PSF}

While we believe the FSC PSF is the most realistic empirical PSF we have been able to produce, we examined the effect of varying the PSF as well.  We explore effects of varying the PSF redshift distribution in Figure 11. Here, we use the PSF computed with three different redshift distributions to analyze the ``all'' simulated images (which were created using the  PSF with the redshift distribution of all 2MIG sources). While the MCMC analysis slightly systematically overestimates the number of counts within 50 kpc for this simulation (see section 5.2), changing to the PSF with the redshift distribution of one of the subsamples of 2MIG galaxies (faint or luminous) also systematically changes the amount of recovered counts. Using the PSF constructed with the redshift distribution of faint galaxies reduces the total inferred number of counts by $\sim 10\%$, while the PSF constructed with the redshift distribution of luminous galaxies increases the total inferred number of counts by $\sim 30\%$. The luminous galaxy PSF  is noticeably flatter than the other PSFs in Figure 5, so this makes sense. In our analysis of the 2MIG data, we match the redshift distributions of the point sources to the distributions of the 2MIG sources, in order to minimize these errors.

\begin{figure}
\plotone{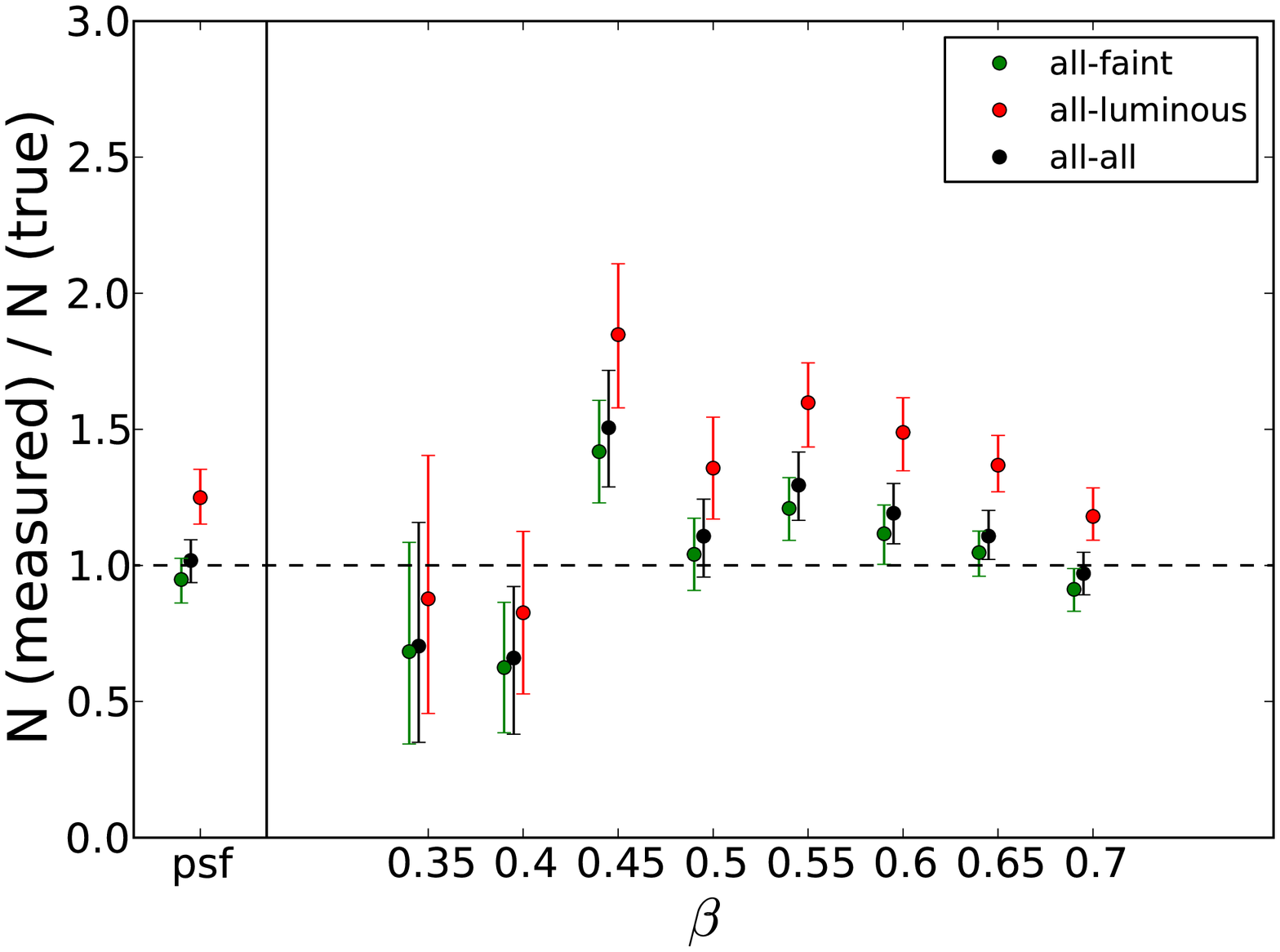}
\caption{ Comparison of pdfs of recovered counts within 50 kpc after variations in the redshift distribution/aperture size of point sources we stack to produce the empirical PSFs. We fit the ``all'' simulated images (created using the redshift distribution of all 2MIG galaxies) with empirical PSFs created using three different redshift distributions as noted in the legend. Using the faint redshift distribution introduces a bias of $\sim 10\%$, while the luminous redshift distribution introduces a bias of $\sim 30\%$ around the fits using the correct relation (the black points), which are themselves systematically high by $\sim$ 20\% - 30\%. }
\end{figure}

\subsection{Including a combination of point source and extended components}
Finally, we examine the effect of mixing an extended component and a point source component in the same image. We created a series of simulated images with half the source counts in the extended component and half in the point source component and analyzed them. Figure 12a shows the total fraction of recovered counts for each set of simulated images, and Figure 12b shows the allocation of the recovered counts into extended and point source components for the set of simulations designed to mimic the full stack of all galaxies. The analysis works very well; the results actually look better than Figures 9 and 10 because the failure to recover very flat extended emission only applies to half of the total source counts, and the point source counts are recovered fully. For the luminous galaxies, the MCMC analysis systematically underestimates the source counts within 50 kpc by $\sim 30\%$; this seems to be due to misattribution of some counts to the extended component, since the number of counts within 500 kpc is recovered correctly. More generally, the division between the two components leads to significant uncertainties on the total counts within each, but in Figure 12b the total number of counts is recovered more accurately than in Figure 9.

\begin{figure*}
\plottwo{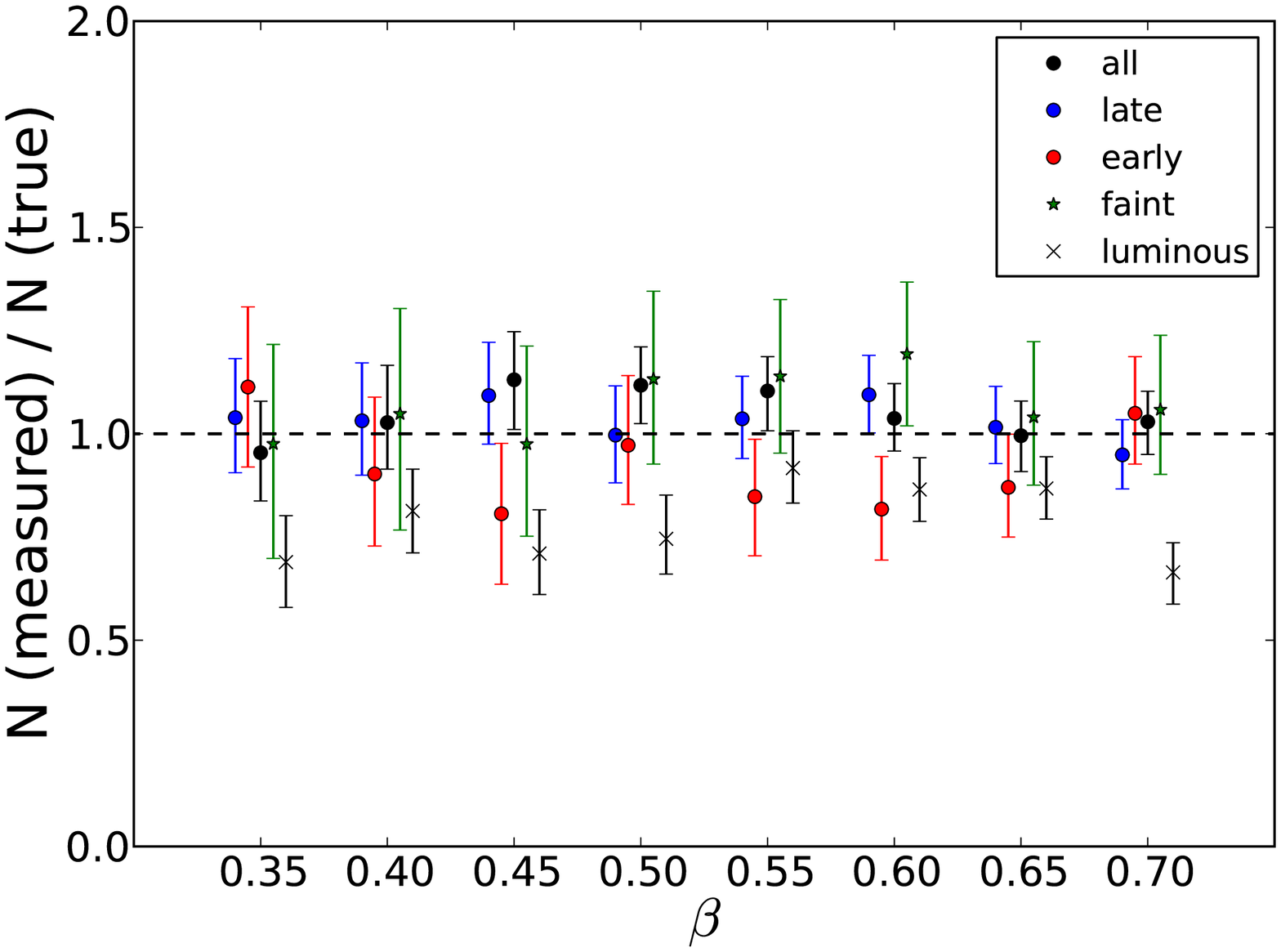}{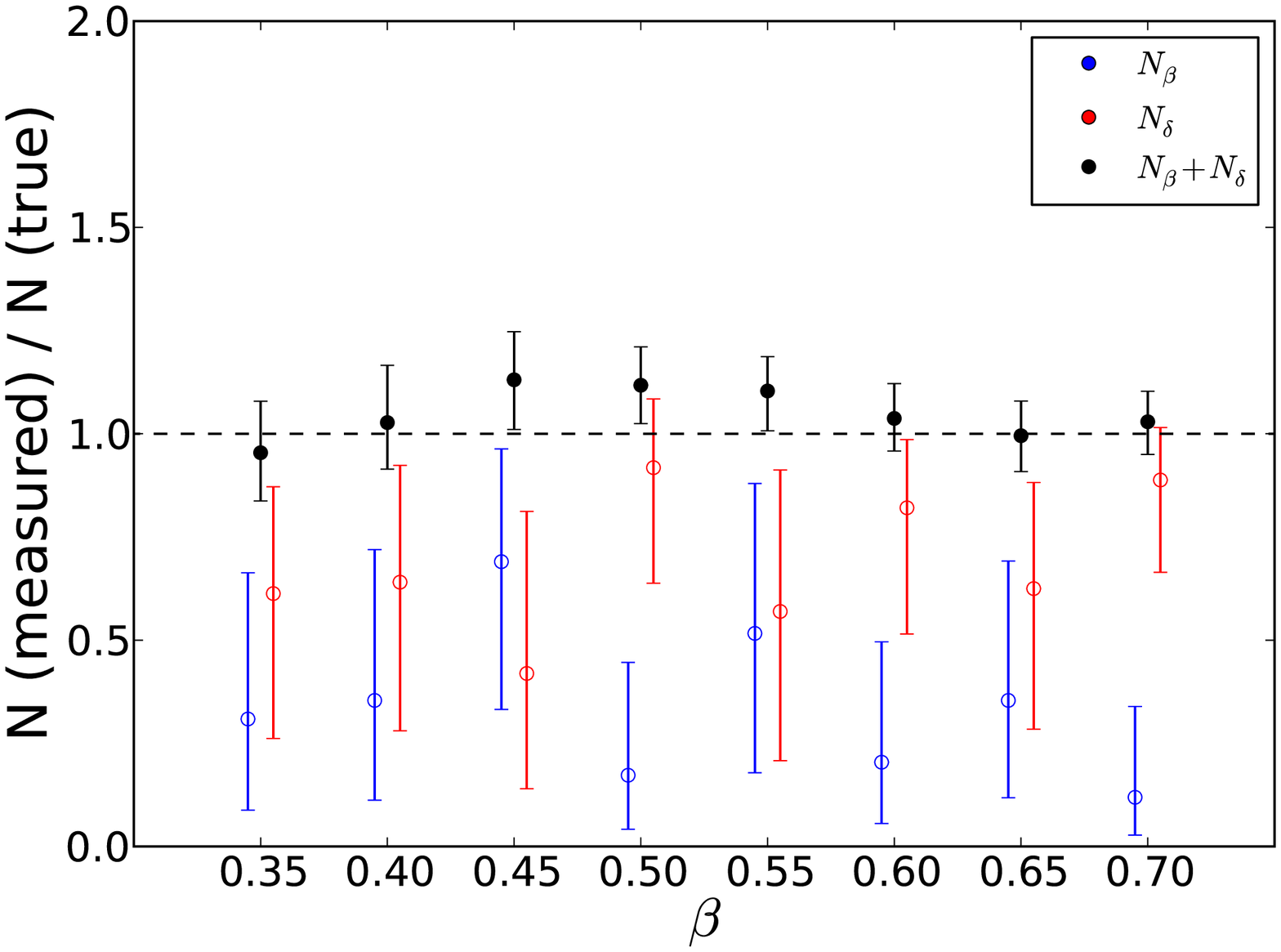}
\caption{ Comparison of pdfs of recovered counts within 50 kpc to the true number of counts within 50 kpc for simulations where $N_{\beta} = N_{\delta}$, i.e. half the source counts are in the extended component and half the source counts are in the point source component. In (a) we present the pdfs of the recovered counts for all five sets of simulations; this Figure shows that the recovered counts are closer to the true value than in the simulations with all the source counts in one component (see Figure 8). In (b), we present the pdfs of the individual components of the ``all'' simulation, and again the recovered values for $N_{\beta}$ and $N_{\delta}$ (both of which should be half the total) are closer to the true value than they are when all the source counts are in just one component (see Figure 9b). }
\end{figure*}

\subsection{Summary of simulations}

The major findings from our simulations are:

1. The MCMC analysis should be able to measure the number of source counts within 50 kpc to an accuracy of 50\% or better. 

2. The MCMC analysis might be able to measure the number of source counts all the way out to 500 kpc, although if $\beta \lapprox 0.5$ it can underestimate the total number of source counts significantly. We will principally be concerned with the emission within 50 kpc for the subsequent analysis.

3. The MCMC analysis cannot reliably recover $\beta$ or $r_0$ with the quality of data we are using.

4. The MCMC analysis has trouble distinguishing purely extended and purely point source emission, especially when focusing on the emission within 50 kpc. Most notably, for $\beta \gapprox 0.5$, it cannot distinguish one component from the other, and attributes most of the emission to the point source component.  

5. If the source is composed of a mixture of an extended component and a point source component, the MCMC analysis can recover the total number of source counts within 50 kpc to an accuracy of 30\% or better. 

6. Reasonable variations in the PSF due to the size of the aperture produce an error in the measured counts within 50 kpc up to 50\%, which we attempt to avoid by choosing the correct redshift distributions when making the empirical PSFs. Reasonable variations in the sample of point sources used to create the empirical PSFs produce a negligible error in the total measured counts.

\section{Statistical significance of detections}

After running the simulations we have confidence in being able to recover the total amount of source emission, especially within a 50 kpc radius of the galaxy, but less confidence in identifying the fraction of emission in the extended component. An important question to consider before running the MCMC analysis is whether we have good reason to believe there {\it is} an extended component in our stacked images, and here we quantify the statistical significance of the detection of extended emission, as well as the statistical significance of the detection of any emission at all. 

To quantify the statistical significance of our detections of {\it any} emission, we apply Poisson statistics on the counts within a radius of 50 projected kpc. We model the background as a uniform component multiplied by the normalized exposure map to account for the ``effective vignetting'' (which is only a few percent within 50 projected kpc). We estimate the level of the uniform background using a bootstrap analysis of various circular annuli around the central region. For each background annulus, we compute the pixel-by-pixel likelihood of obtaining the observed counts within 50 projected kpc given the background annulus and vignetting model. We compare this likelihood to 10000 randomly selected regions of radius 50 projected kpc chosen across the stacked image (but outside the central region). We repeat this analysis for 100 variations in background annuli, yielding a total of $10^6$ null likelihoods. The statistical significance of our source detection is the fraction of null likelihoods which are lower than the observed likelihood of the central emission. These values are listed in Table 3. 

We can also approximate the significance of the detection as the conjugate of the probability of the measured number of counts within 50 projected kpc, given our estimate of the background count rate from the MCMC fitting. This method is simpler, but does not account for non-uniformity in the background, and therefore may overestimate the significance of the detections if the background has significant variations above Poisson noise. Therefore we only use this technique to approximate the significance of detections at confidences of greater than $0.999999$ (the highest confidence we can measure from the bootstrap technique above). These values are listed in Table 3 for the stacks of all galaxies, early-type galaxies, luminous galaxies, and luminous early-type galaxies. 

To quantify the statistical significance of our detections of extended emission, we employ a likelihood ratio test. We compare the likelihood of the data for a model with an extended component (eq. 3) to the likelihood of the data for a model with the extended component set equal to zero (eq 3 with $A_{\beta} = 0$). The latter model therefore only contains a uniform background and a point source convolved with the empirical PSF (both terms subject to effective vignetting as well). We use a truncated Newton algorithm\footnote{the tnc algorithm included in the Python package SciPy.optimize, v. 0.11} to find the maximum likelihood best-fit to the data for each of our two models. In the model with the nonzero extended component, we require $r_0$ to be between 0.5 and 5.0 kpc and we require $\beta$ to be between 0.35 and 0.70. We then compute the likelihood ratio for the best fits, defined as the difference between the log-likelihoods of the best fits for the two models. Since the model without an extended component is a subset of the model with the extended component, the likelihood never decreases when we allow for an extended component, so the likelihood ratio is always non-negative. 

To interpret the likelihood ratios, we compare the likelihood ratios for the data to likelihood ratios computed for simulated data. For each stacked image, we generate 1000 simulated images using the parameters of the simulations we ran in section 5. These simulated images have all the source counts in the point source component, however. We find the maximum-likelihood fits to each image for both models, and compute the likelihood ratio for each simulated image. This gives us a set of 1000 likelihood ratios for images without an extended component, and we therefore can estimate the probability of measuring a given likelihood ratio for an image without any extended emission. The statistical significance of our detection of an extended component is then just the conjugate of this probability, i.e. the fraction of point source simulations with likelihood ratios lower than the value we measure for the data. These values are listed in Table 3.

\begin{deluxetable*}{ccc}  
\tablecolumns{3}
\tablecaption{Significance of detections}
\tablehead{   
  \colhead{sample} &
  \colhead{$1 - p$(any emission $r<50$ kpc)} &
  \colhead{$1 - p$(extended emission)}
}
\startdata
All galaxies & $4.4\times10^{-16}$ $(8.1 \sigma)$ &    0.073 \\
Late-type galaxies& 0.000446  & 0.347\\
Early-type galaxies& $ 3.0\times10^{-9}$ $(5.9 \sigma)$ & 0.107 \\
Faint galaxies & 0.005312 & 0.528 \\
Luminous galaxies & $2.6\times10^{-23}$ $(9.9 \sigma)$ &0.007\\
Luminous late-type galaxies&  0.000025 & $0.013$\\
Luminous early-type galaxies& $ 3.2\times10^{-15}$ $(7.9 \sigma)$& $0.079$\\
Random locations & 0.412603  & -
\enddata
\tablecomments{Estimated statistical significance of our detections of: any emission above the background (column 2), and extended emission (column 3). Significances expressed in scientific notation are higher than $99.9999\%$ and therefore estimated using a different, less precise, technique. See section 6 for details of how these significances were computed. The image of ``random locations'' can be seen in Figure 2; we stacked images of 2500 random locations in the sky to serve as a null hypothesis. Because emission is not detected, we did not compute the significance of extended emission for this image. }
\end{deluxetable*}

We computed the significance of emission in the full stack and the four subsamples used throughout the analysis, but we also examine here a few other interesting subsamples. We stacked the 556 luminous late-type galaxies and the 355 luminous early-type galaxies separately. There are not enough photons in these images to get as reliable a measurement of the shape of the extended component for these stacks, so we do not emphasize them in much of the subsequent analysis, but the background is so low in these images that we can detect the emission at very high confidence, so we include these figures in Table 3. We also include the stack of random pointings in the sky shown in Figure 2, as a null hypothesis, which indeed is not detected at any significance. 

Except for the faint galaxies, all of the samples are detected at $> 3\sigma$ confidence (and in some cases, the signal is much stronger). None of the extended emission is detected at $>3\sigma$ confidence, though the luminous galaxies and the luminous late-type galaxies are close to $3\sigma$. This latter result appears to be a fundamental limitation of the \emph{ROSAT} data. After all, we only have approximately one source photon per luminous galaxy in our stacked images, and less than one source photon per faint galaxy. Most of our samples have extended emission detected above or near 90\% confidence (the full stack, the early-type galaxies, the luminous galaxies, the luminous late-type galaxies, and the luminous early-type galaxies), while two of our samples (the late-type galaxies and the faint galaxies) have extended emission at much lower confidence. As we will show in section 7.3.1, these samples also have stronger evidence for hot gas in their spectral hardness ratios than the other two samples, and the preponderance of evidence will suggest (but not definitively) that these galaxies do indeed have extended gaseous emission. However, with \emph{ROSAT}, the result is still only a strong suggestion of extended emission; a definitive detection will probably have to wait for the next generation X-ray all sky survey (i.e E-Rosita).  \\

\section{Results}
Based on our simulations,  we have confidence in being able to use MCMC analysis to extract the total number of source counts from our stacked 2MIG data. The simulations  suggest that the MCMC analysis is less reliable in recovering the fraction of source counts in the extended component, although it is fairly reliable in the (likely) case of an equal mixture of point source emission and extended emission. We also have very strong (up to nearly $10 \sigma$) detections of soft X-ray emission from all of our stacked samples, and moderate confidence that there is extended emission around the stacks of all galaxies (92.7\% significance), early-type galaxies (89.3\% significancw), luminous galaxies (99.3\% significance), luminous late-type galaxies (98.7\% significance), and luminous early-type galaxies (92.1\% significance). 

We now present the MCMC analysis on the real stacked images. Just as with the simulations, we specify the appropriate PSF and the stacked exposure map for each image, as well as prior distributions for the five model parameters in eq. 3.  The prior for $A_{\delta}$ was uniformly distributed between 0 and $2\times10^4$, and the prior for $A_{\beta}$ was uniformly distributed between 0 and $2\times10^5$\footnote{Note that $A_{\delta}$ translates directly into the number of point source counts in the image, while the number of extended counts depends on $A_{\beta}$, $r_0$, and $\beta$. The adopted priors on $A_{\beta}$ and $A_{\delta}$ correspond to roughly equal ranges in counts for each component}. $A_c$ was normally distributed around the average pixel value at large radii, after attempting to correct for the ``effective vignetting'', with a standard deviation of $0.1$. The core radius $r_0$ was uniformly distributed between 0.5 and 5.0 kpc, and $\beta$ was normally distributed around $\beta = 0.5$, with a standard deviation of $0.1$. Note that our priors for $\beta$ and $r_0$ reflect our prior knowledge of the shapes of hot gaseous halos around other galaxies, including the two giant spirals NGC 1961 and UGC 12591. 

We carefully explored the effects of various choices for the prior distributions of the free parameters listed above. Varying the width of the priors for $A_{\delta}$ and $A_{\beta}$ does not affect the result significantly, since the total amount of emission above the background is fairly well constrained by the data. Similarly, small variations in the width or the shape of the prior on $A_c$ do not have much effect, since the background is also well constrained by the data. Varying $r_0$ has almost no effect on the luminosity and only a small effect on the mass, but the simulations showed that the data are not of sufficient quality to constrain $r_0$ at all so we leave the priors wide. Finally, varying $\beta$ does have a significant effect on the posterior mass distribution; however, we think we understand $\beta$ well, and nearly all existing observations of $\beta$ around other galaxies place it between $0.4$ and $0.6$, with a median right around $0.5$. Priors on $\beta$ that conform to these constraints all yield similar results. 

Figure 13 shows the pdf of the total number of source counts within physical radii of 50 kpc and 500 kpc from the central galaxies, for each of the four subsamples of the 2MIG galaxies. The number of counts within 50 kpc is fairly well constrained, but the number of counts within 500 kpc has a long tail extending to much larger values. This tail reflects the possibility of flattened profiles, as examined in our simulated images. 1-$\sigma$ confidence intervals miss this long tail, but it appears plainly in the full marginalized probability distribution.

\begin{figure*}
\plottwo{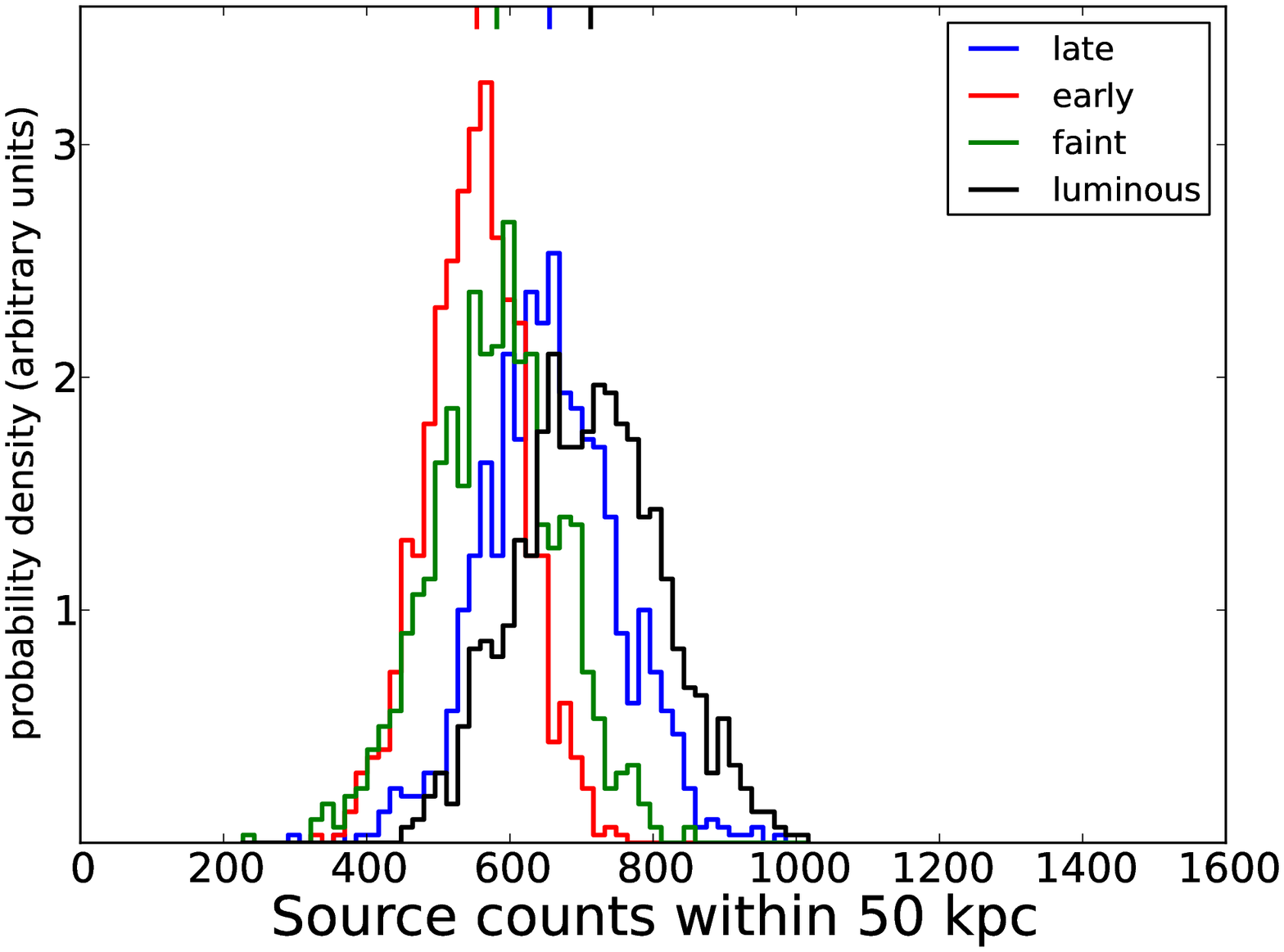}{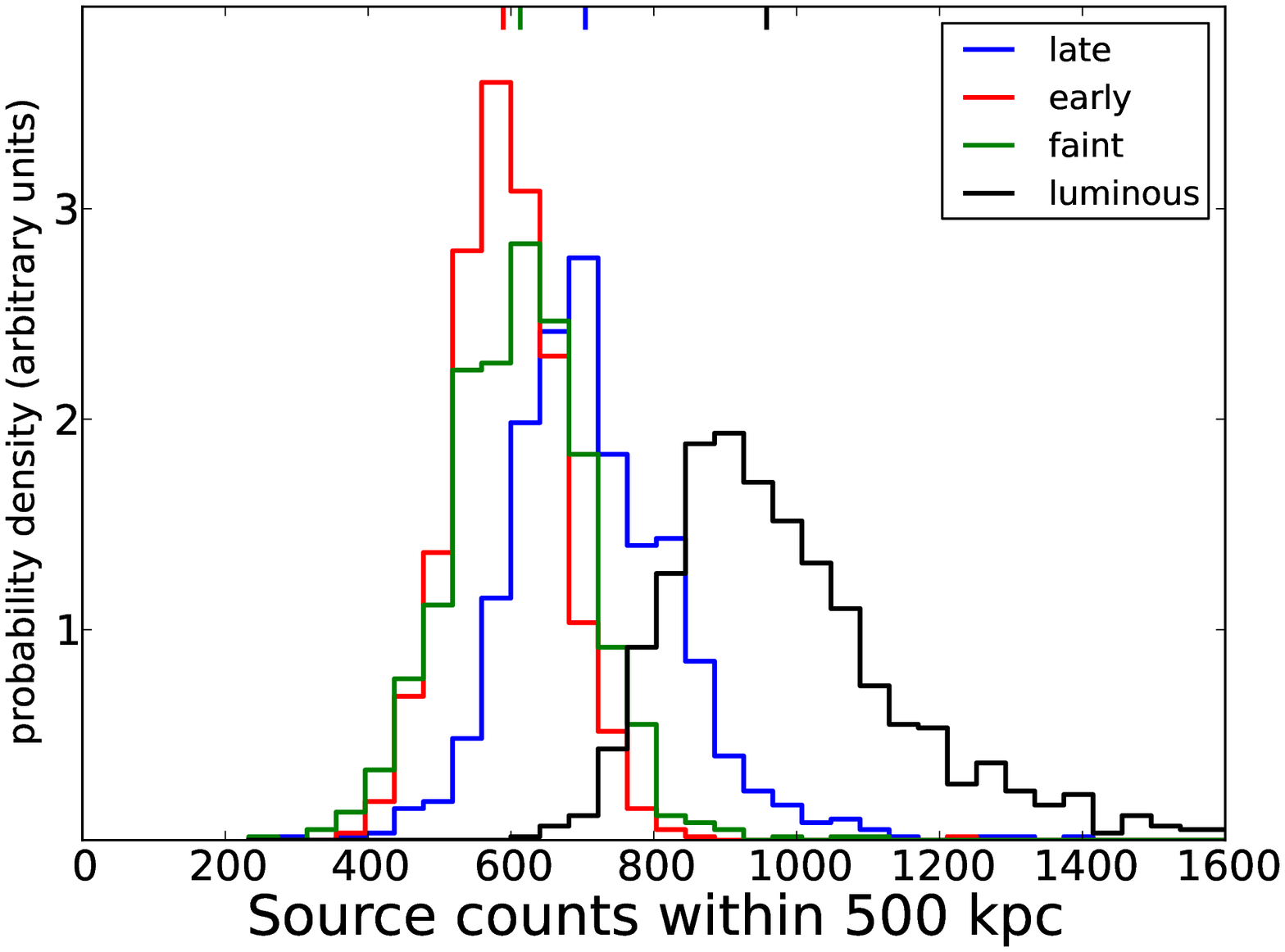}
\caption{ Probability distribution functions (pdfs) for the total number of source counts in the stacked 2MIG images, within a radius of (a) 50 kpc and (b) 500 kpc from the central galaxies. The four samples are shown in four different colors as indicated, and the median of each pdf is marked with a vertical dash at the top of each plot. The longer tails to the right in (b) reflect the possibility of extremely flattened profiles. }
\end{figure*} 

\subsection{Average Luminosity}

The next step is to convert the number of source counts in each image into an average luminosity per galaxy. This requires three steps. First, we must assume a spectral model for the soft X-ray emission, and use the spectral model to convert the number of counts into units of energy (erg). At the same time we can fold in the instrumental effective area as a function of energy to get energy per unit area (erg cm$^{-2}$). Then we divide by the exposure time, which we estimate from the stacked and weighted exposure map, to get a flux (erg cm$^{-2}$ s$^{-1}$). Finally, we multiply by $4 \pi d_L^2$ to produce a luminosity (erg s$^{-1}$), which is simple because the exposure map was weighted such that each galaxy's exposure time corresponds to flux from at an effective distance of 100 Mpc (see section 3). 

There is also a small additional dependence on redshift, which changes the effective energy band of our observations by a factor of $(1+z)$ and introduces cosmological surface brightness dimming which goes as $(1+z)^4$. The maximum redshift in our sample of $0.06$ and the mean is $0.02$. The galaxies which dominate the observed flux have even lower redshifts than that, so these redshift effects are small and we do not attempt to correct for them. 

For the assumed spectral model, we examine various linear combinations of an APEC model (\citet{Smith2001}; used for the hot gas) and a powerlaw (for the X-ray binaries / possible AGN), as well as various amounts of absorption for both components. For the APEC model, we vary the metallicity between $0.2 Z_{\odot}$ and $1.0 Z_{\odot}$, and we explore three different gas temperatures. The $kT$ = 0.2 keV model is approximately the virial temperature for hot gas with a truncated singular isothermal sphere density profile inhabiting an $L*$ galaxy's halo, and represents an approximately quasi-static hot gas halo. We also explored the effect of hotter gas (0.3 keV) and cooler gas (0.1 keV). For the X-ray binaries, we assumed a powerlaw with spectral slope of $1.56$ \citep{Irwin2003}, and allowed the slope to vary by $\pm 0.2$. For the AGN, we assumed a spectral slope of $2 \pm 0.5$. Finally, for each model we vary the absorbing Hydrogen column between $10^{20}$ cm$^{-2}$ and $10^{21}$ cm$^{-2}$, a range which encompasses 80\% of our 2MIG galaxies based on the \citet{Dickey1990} Galactic values.  

We compute the conversion factors for each model using the HEASARC PIMMS tool\footnote{http://heasarc.nasa.gov/Tools/w3pimms.html} (version 4.6). In addition to calculating the spectral models, this tool automatically folds in the effective area for the instrument, so the output is in units of erg cm$^{-2}$. We will estimate the conversion factor from counts to energy per area by computing the linear combination of these models. We attribute the AGN-like spectrum to the point source emission, and the combination of hot gas and X-ray binaries to the extended component. The point source being an AGN is plausible both based on the subsequent point source luminosities and a recent optical emission line analysis of isolated galaxies that found that nearly all of these galaxies harbor a weak AGN \citep{Hernandez-Ibarra2012}. Since the X-ray binary spectrum ($\Gamma=1.56$) is within the range of possible AGN spectra we consider ($\Gamma = 1.5-2.5$), this also allows for the possibility of X-ray binaries in the central region of the galaxy. We can then estimate the appropriate linear combination of these components based on the fraction of emission in each spatial component, as measured by our MCMC analysis. This fraction is shown for each sample in Figure 14. 

\begin{figure}
\plotone{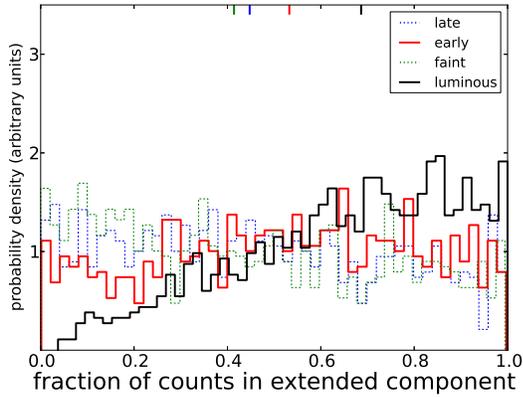}
\caption{ The pdfs for the fraction of source counts within 50 kpc of the galaxy, attributed to the extended component for all four subsamples. The median of each distribution is marked with a vertical dash at the top of the plot. The pdfs for late-type and faint subsamples are shown as dashed lines because the statistical significance of the extended component in these samples is much lower than for the others. For all four samples the uncertainties are very large, but the median extended fractions increase with the statistical significance of the extended component, up to a maximum of $\sim 70\%$ for the luminous galaxies. }
\end{figure} 

The median extended fractions increase with the statistical significance of the extended component, up to a maximum of $\sim 70\%$ for the luminous galaxies, though the uncertainties are very large and these images are all also consistent with an equal mixture of extended and point source emission (i.e. an extended fraction of 50\%). In order to estimate the counts-to-energy ratio, we use the fraction of extended emission for the  full sample instead of the subsamples, since they are all statistically indistinguishable from one another, but the full sample has the most photons and theoretically the most accurate constraints. This fraction is $60\pm30$\%. We assign the remaining fraction of the emission to the point source component, and assume it has the AGN-like spectrum described above. For the extended component, we explore models with 100\% hot gas, 100\% XRB, and 50\% of each. We present the computed counts-to-energy ratios in Figures 15a and 15b. 

\begin{figure*}
\plottwo{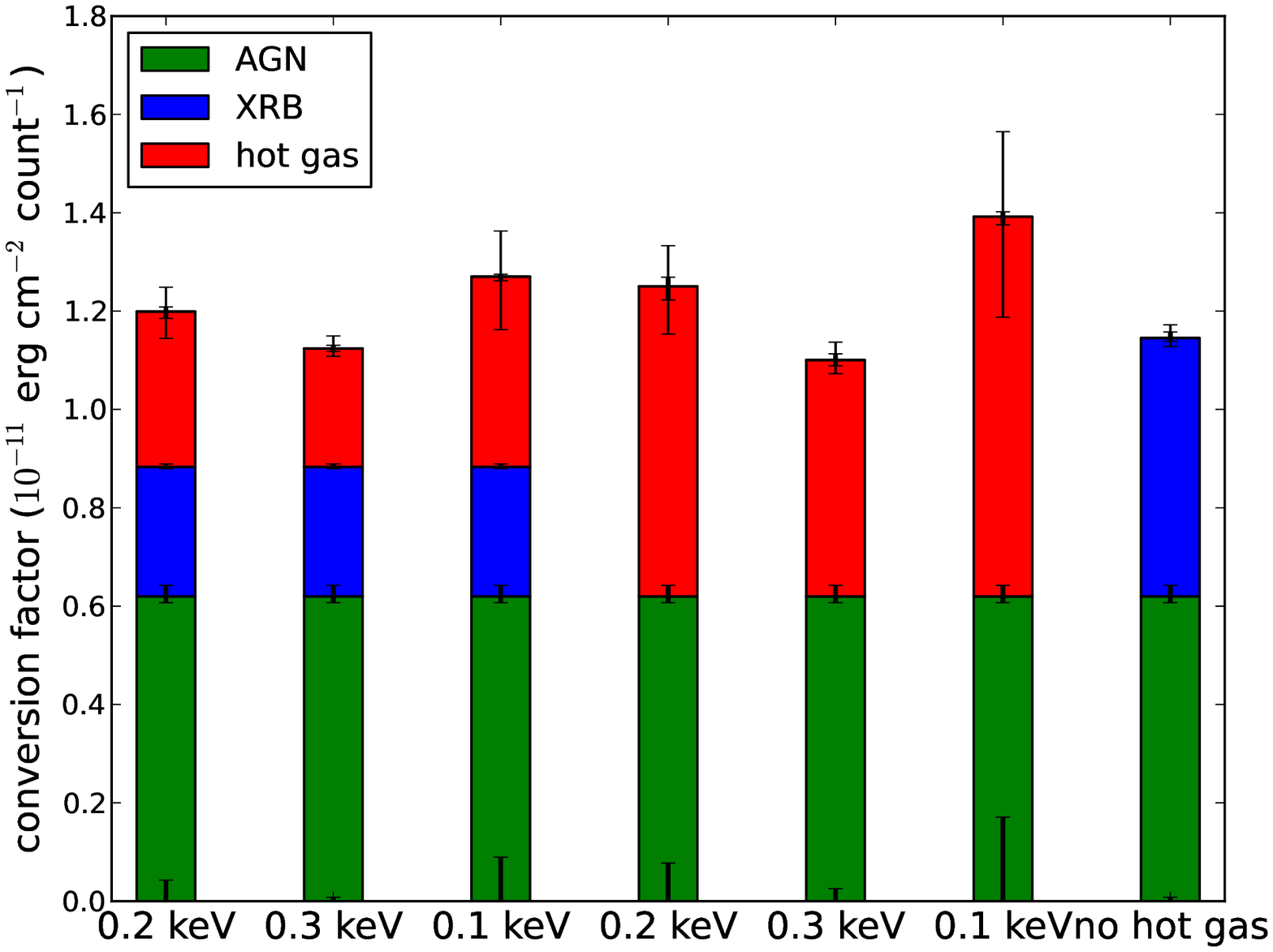}{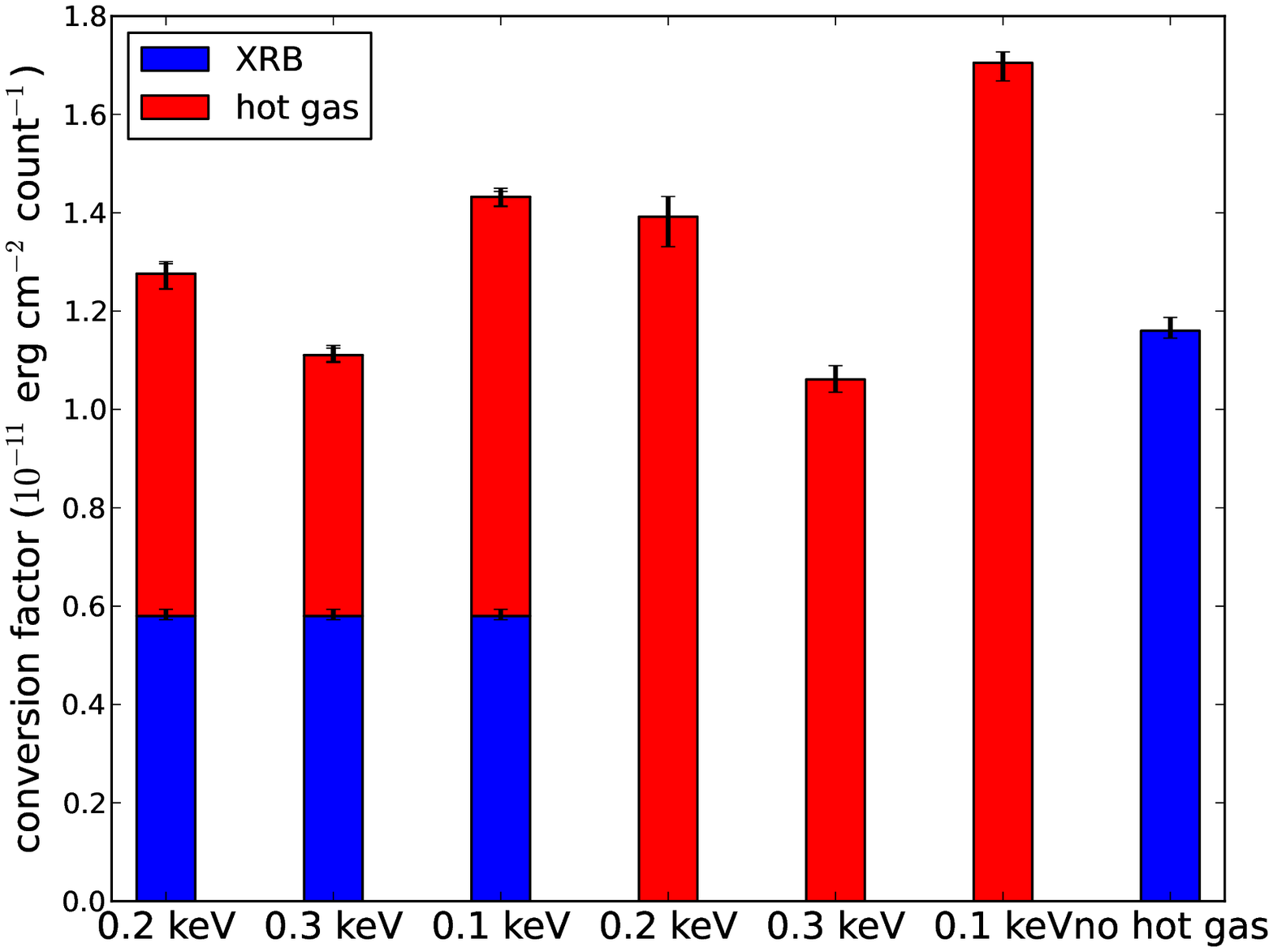}
\caption{ Estimates of the conversion factor from counts to energy per unit area for our data, assuming various spectral models for the emission. In (a) we consider models with an AGN component, an X-ray binary component, and a hot gas (APEC) component - i.e. a point source and extended emission. In (b) we only consider the latter two components - i.e. extended emission only. The assumed temperature of the hot gas is noted at the bottom of each model. The error bars at the top of each component show the uncertainties contributed to the total conversion factor by uncertainties in the shape of the spectral model for that component. The error bars at the very bottom of each model show the uncertainty in the total conversion factor from the uncertainties in the fraction of AGN emission. The largest error bar at the top of each model shows the sum of all the uncertainties, added in quadrature. We adopt the leftmost conversion factors from each plot for the rest of the analysis - $1.21\times10^{-11}$ erg cm$^{-2}$ count$^{-1}$  and $1.28\times10^{-11}$ erg cm$^{-2}$ count$^{-1}$ for the total emission and the extended emission, respectively.  }
\end{figure*}

We think the most plausible model is the leftmost model in each plot, so we adopt conversion factors of $1.21\times10^{-11}$ erg cm$^{-2}$ count$^{-1}$  and $1.28\times10^{-11}$ erg cm$^{-2}$ count$^{-1}$ for the total emission and the extended emission, respectively. The uncertainty in these conversion factors can be estimated from the range of possible values visible in Figures 15a and 15b, and seems to be no more than 30\% upwards and no more than 15\% downwards. Most of this upwards scatter comes from the models with the extended component consisting entirely of hot gas at 0.1 keV, and we will show later (section 7.3) that these models are unlikely to describe the real data, since X-ray binaries almost certainly contribute significantly to the observed emission.

For the total effective exposure, as discussed above in section 3, we already have a stacked exposure map for each sample, where the exposure map of each image is weighted by the quantity $(100\text{ Mpc } / d)^2$ so that the effective distance of the sample is 100 Mpc. These exposure maps peak at the center of the image and decline slowly at larger radii (by about 10\% or so) because not every image in the stack extends out to 500 projected kpc in every direction. This ``effective vignetting'' is included in our model for the data, so the appropriate exposure time to use for converting energy/area to flux is the value at the center of the image. Thus the the stack of late-type galaxies has an effective exposure time of $1.51\times10^6$ s, the early-type galaxies have an effective exposure of $5.63\times10^5$ s, the faint galaxies have an effective exposure of $1.71\times10^6$ s, and the luminous galaxies have an effective exposure of $3.65\times10^5$ s. 

We can now combine the counts-to-energy factor, the distance, and the effective exposure time to convert the counts probability distribution function into a luminosity probability distribution function (Figure 16).  The luminosity within a radius of 500 kpc is not a physically interesting quantity, so we show the luminosity within 300 kpc instead, as this is closer to the virial radius of an $L*$ galaxy.  There is very little emission beyond 50 kpc, however, so the choice for the outer radius is not very important.

\begin{figure*}
\plottwo{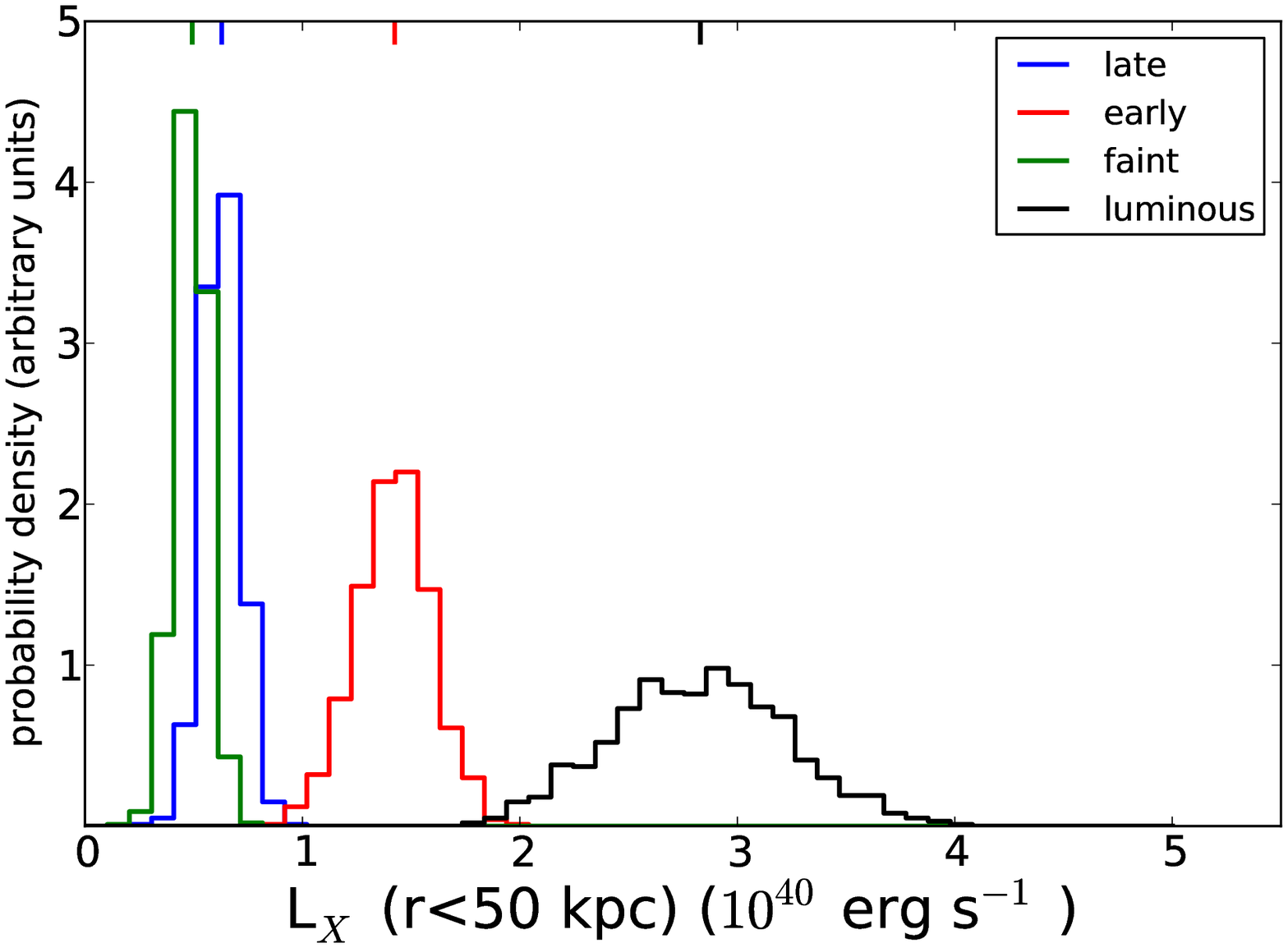}{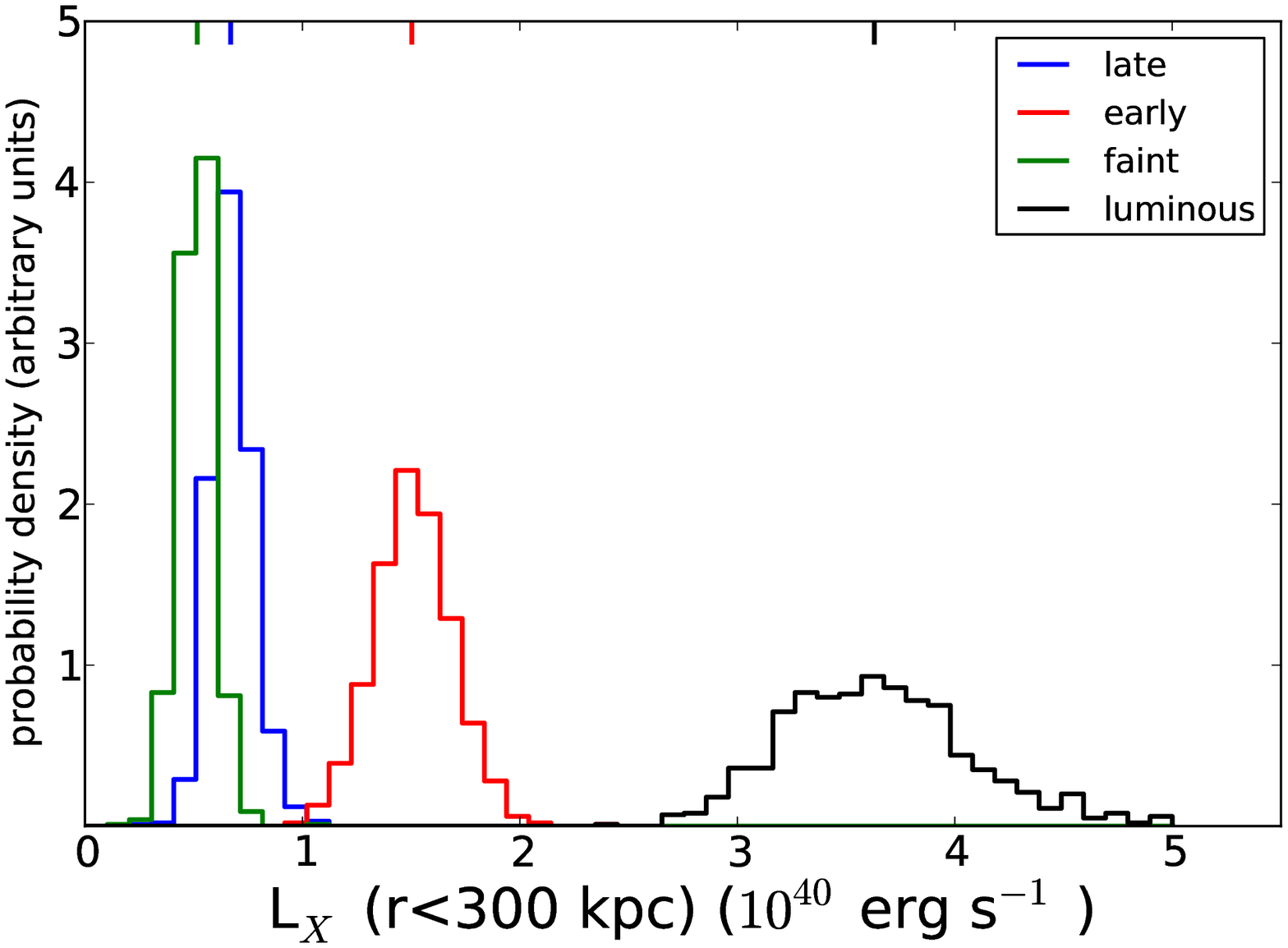}
\caption{ Probability distribution functions (pdfs) for the average 0.5--2.0 keV X-ray luminosity within radii of (a) 50 kpc and (b) 300 kpc of the galaxy, for each of our four subsamples of 2MIG galaxies. The median of each distribution is marked with a vertical dash at the top of the plot. Note that the X-axis can be slightly shifted left or right due to uncertainties in the conversion from source counts to luminosity, as illustrated in Figure 15.  }
\end{figure*} 

Finally, from the MCMC analysis, we can separate out just the extended component of the emission by integrating only the emission parameterized by the $\beta$-model. The average luminosity from this extended emission is shown in Figure 17. 

\begin{figure*}
\plottwo{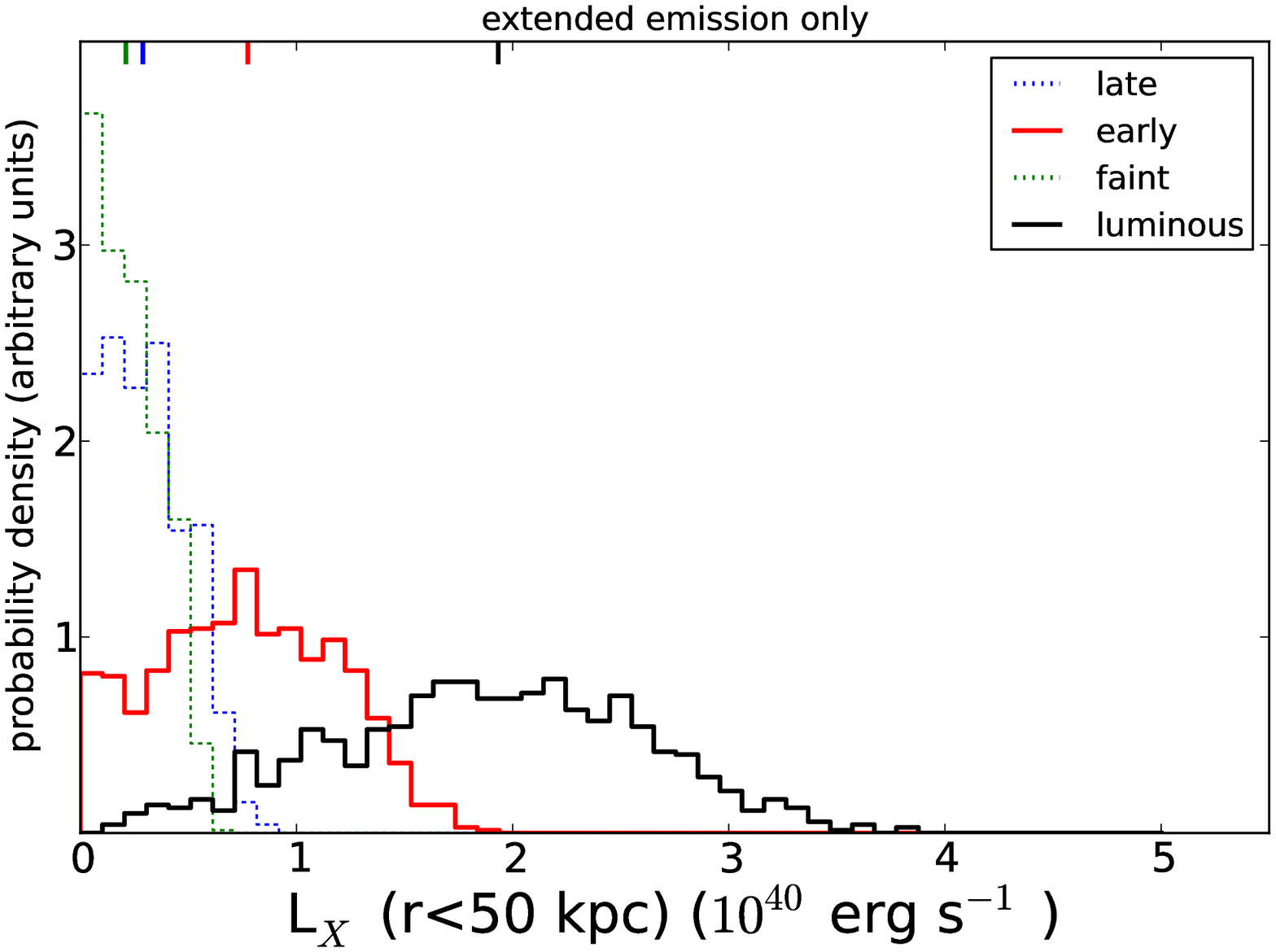}{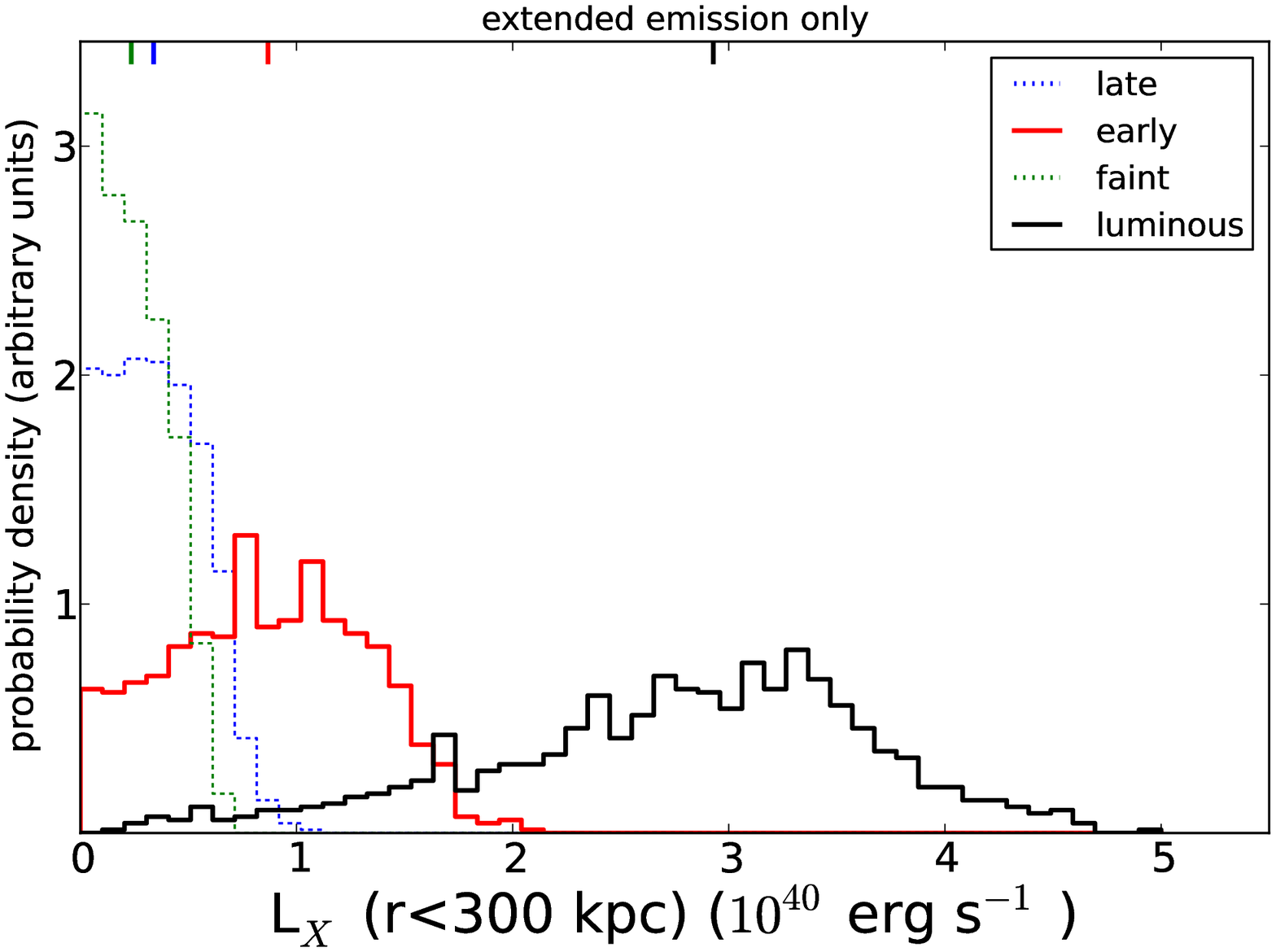}
\caption{  Probability distribution functions (pdfs) for the average 0.5-2.0 keV X-ray luminosity of the {\it extended component} of the emission within radii of (a) 50 kpc and (b) 300 kpc of the galaxy for each of our four subsamples of 2MIG galaxies. The median of each distribution is marked with a vertical dash at the top of the plot. The pdfs for the late-type and faint subsamples are shown as dashed lines because the statistical significance of the extended component is well below 90\%. Note that the X-axis can be slightly shifted left or right due to uncertainties in the conversion from source counts to luminosity, as illustrated in Figure 15.   }
\end{figure*}

\subsection{AGN emission}

While the inclusion of an AGN component does not change the estimation of the total luminosity significantly, it is worth reflecting on the expected luminosity of accretion onto supermassive black holes in these galaxies. Based on the spatial analysis, we are placing about 50\% of the emission into the point source component (which includes both AGN emission and X-ray binaries in the central 5 kpc of the galaxy). In detail, the average soft X-ray luminosities for the point source component are $3\times10^{39}$ erg s$^{-1}$, $7\times10^{39}$ erg s$^{-1}$, $3\times10^{39}$ erg s$^{-1}$, and $9\times10^{39}$ erg s$^{-1}$, for the late, early, faint, and luminous samples, respectively. We know there are no AGN accreting at luminosities of $10^{41.5}$ erg s$^{-1}$ or more in our galaxies, because we examined each image and removed any images with extremely bright point sources, as well as automatically removing any objects in the \emph{ROSAT} BSC. But we also expect most of these galaxies to harbor at least some SMBH accretion, based on high excitation lines visible in optical spectra of isolated nearby galaxies  \citep{Hernandez-Ibarra2012}. 

To study the effect of AGN emission more quantitatively, we examined the AMUSE-field sample \citep{Miller2012}, a Chandra survey of nearby early-type field galaxies designed to characterize the intensity of low-luminosity AGN activity in these galaxies. We divided the 57 AMUSE-field galaxies in the lowest-density environments into two stellar mass bins, log $M*$ = 9.5-10.5 (23 galaxies, roughly corresponding to the range covered by our ``faint'' sample) and log $M*$ = 10.5-11.5 (34 galaxies, roughly corresponding to the range covered by our ``luminous'' sample). The faint galaxies all had AGNs fainter than $10^{39.5}$ erg s$^{-1}$, and only 17\% of them even had emission at the $10^{39}$ erg s$^{-1}$ level. The luminous galaxies all had AGNs fainter than $10^{41.1}$ erg s$^{-1}$, 41\% of them had emission below $10^{39}$ erg s$^{-1}$, and the remaining 59\% had a median luminosity just under $10^{39.5}$ erg s$^{-1}$. Both of these samples are fully consistent with our average point source luminosities, and in fact our point source luminosities are about twice the average measured SMBH luminosities for the AMUSE-field galaxies. The other half of the point source component can be attributed to X-ray binaries in the central 5 kpc of the 2MIG galaxies.

\subsection{X-ray binaries}

It is therefore also worth examining the global effect of X-ray binary contamination. They are the most important source of soft X-rays in galaxies other than hot gas and AGN (e.g. \citealt{Boroson2011}), and they can extend out to tens of kpc. We constrain X-ray binary emission using two independent methods, detailed below.

\subsubsection{Hardness Ratio}

To attempt to distinguish between hot gas emission and X-ray binary emission, we stacked each sample of 2MIG galaxies in a ``soft'' 0.5-1.0 keV band and a ``hard'' 1.0-2.0 keV band, and fit for the total amount of emission within 50 kpc above the background. Ideally we would examine just the extended component, but we do not have enough counts in these narrower bands to constrain the extended component significantly. We computed the hardness ratio (HR), defined as 

\begin{equation}HR \equiv \frac{ \text{H} - \text{S}}{\text{H} + \text{S}} \end{equation}

where H and S are the number of counts in the hard and soft components, respectively. The 68\% confidence intervals for HR for the four subsamples are shown in Figure 18. In this Figure we also plot the expected HR for various emission models. The green lines are powerlaw emission models with a slope of $-2$ and absorbing columns of 1,5, and 10 $\times 10^{20}$ cm$^{-2}$. These are intended to mimic AGN emission, although as discussed above it is difficult to distinguish AGN and XRB emission based on their spectral hardness in the soft X-rays. The blue lines are powerlaw emission models with a slope of $-1.56$ and absorbing columns of 1,5, and 10 $\times 10^{20}$ cm$^{-2}$, designed to mimic XRB emission. And the three sets of red lines are APEC emission models with $kT$ = 0.1, 0.2, and 0.3 keV, respectively. For the APEC models, the metallicity ranges from $0.2 Z_{\odot}$ to $1.0 Z_{\odot}$ and the absorbing columns range from $10^{20}$ to $10^{21}$ cm$^{-2}$. Higher absorption increases the hardness ratio. 

\begin{figure}
\plotone{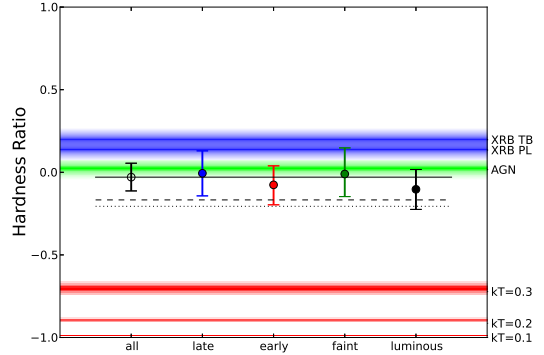}
\caption{ Hardness ratio of source counts within 50 kpc of the galaxy, for the full sample and all four samples of 2MIG galaxies, along with hardness ratios predicted by various spectral models. The shaded red region corresponds to APEC models for hot gas emission at the indicated temperatures; the region encompasses different choices for metallicity (ranging from $0.2 Z_{\odot}$ to $1.0 Z_{\odot}$ and absorption (ranging from $1\times10^{20}$ cm$^{-2}$ to $1\times10^{21}$ cm$^{-2}$), with the darkest shading at the central values for these choices. The shaded green region corresponds to a model for AGN emission (an absorbed $\Gamma = 2$ powerlaw), and the blue shaded region corresponds to the union of two different models for XRB emission; note that the green and blue regions overlap significantly. For full details of these models, see the text. In all models, increasing absorption increases the hardness ratio. The data primarily fall between the XRB and APEC models, and overlap with the AGN models. The fact that no linear combination of XRB + AGN model can reproduce the observed hardness ratios suggests that an APEC (hot gas) component is required as well. The three black lines correspond to three linear combinations of the three components. The dotted black line is an equal combination of all three components, the dashed black line is the leftmost model from Figure 15a with approximately a 2:1:1 ratio of AGN : APEC : XRB emission. The solid black line has 1/4 AGN emission and finds the best fit for the APEC and XRB fractions in the full sample of 2MIG galaxies: 60\% and 15\%. }
\end{figure} 

We can see that all five samples have statistically indistinguishable hardness ratios. In general no linear combination of AGN and XRB can reproduce the observed hardness ratios, except pure AGN absorption, and that is disfavored by both the spatial analysis and the expected average AGN luminosity. If we include any XRB emission, we need to include APEC emission as well to keep the total hardness ratio soft enough to fit the data. Thus, since we expect XRB emission in each sample, the measured hardness ratios imply the presence of hot gas in each of our samples as well.

We can make a simple estimate of the XRB and APEC contributions. We adopt what we consider the likeliest values for each spectral model -  the XRB and AGN models with $N_H = 5\times10^{20}$ cm$^{-2}$, and the APEC model with $kT = 0.2$ keV, $Z = 0.5 Z_{\odot}$, and $N_H = 5\times10^{20}$ cm$^{-2}$. We attribute 1/4 of the emission to SMBH accretion (with an absorbed $\Gamma = 2$ powerlaw), and the rest to XRBs and the hot gas. Since the spatial analysis suggests $\approx 1/2$ of the emission belongs in the point source component, this implies the rest of the point source emission comes from X-ray binaries. Solving for the resulting fraction of the total emission attributed to the other components, we find $f_{\text{XRB}} = 60$\% and $f_{\text{APEC}} = 15$\%.

\subsubsection{Scaling Relations}

We reach a similar conclusion if we estimate the X-ray binary luminosity from established scaling relations. There is a correlation between $L_K$ for a galaxy and its low-mass XRB (LMXB) luminosity. The most recent estimate \citep{Boroson2011} finds $L_{\text{LMXB}} \approx 1 \times 10^{29} (L_K / L_{\odot}) $ erg s$^{-1}$. This is a broad-band measurement of $L_X$ (0.3-8 keV); we use their spectral model (thermal bremsstrahlung with $kT = 7$ keV, although they note a powerlaw with $\Gamma = 1.4-1.8$ works about as well) to convert this $L_X$ into our 0.5-2.0 keV band. We assume $N_H = 5 \times10^{20}$ cm$^{-2}$. The resulting scaling relation is

\begin{equation}L_{\text{LMXB}} \approx 3\times10^{28} \frac{L_K}{L_{\odot}} \text{ erg s}^{-1} \end{equation}

For high-mass X-ray binaries, the luminosity scales with the galactic star formation rate (SFR). The most recent estimate \citep{Mineo2012} is $L_{\text{HMXB}} \approx 2.6 \times 10^{39}$ SFR erg s$^{-1}$ for a SFR in $M_{\odot}$ yr$^{-1}$. Again, this is a broad-band measurement of $L_X$ (0.5-8 keV), and we use their spectral model (a $\Gamma = 2$ powerlaw) with $N_H = 5 \times10^{20}$ cm$^{-2}$ to convert into our 0.5-2.0 keV band. The resulting scaling relation is

\begin{equation}L_{\text{HMXB}} \approx 1.4\times10^{39} \frac{\text{SFR}}{M_{\odot} \text{ yr}^{-1}} \text{ erg s}^{-1}\end{equation}

We can use these scaling relations to estimate the expected XRB luminosity for each of our subsamples.  We compute the mean exposure-weighted $L_K$ for the late-type, early-type, faint, and luminous subsamples: $4.4\times10^{10} L_{\odot}$, $7.4\times10^{10} L_{\odot}$, $3.3\times10^{10} L_{\odot}$, and $1.4\times10^{11} L_{\odot}$, respectively. This yields $L_{\text{LMXB}}$ of $1.3\times10^{39}$ erg s$^{-1}$, $2.2\times10^{39}$ erg s$^{-1}$, $1.0\times10^{39}$ erg s$^{-1}$, $4.3\times10^{39}$ erg s$^{-1}$, respectively. We assume mean SFRs of $0.1 M_{\odot}$ yr$^{-1}$ for the early-type galaxies (including S0s) and $1.0 M_{\odot}$ yr$^{-1}$ for the late-type galaxies, and since about 2/3 of the galaxies are late-type, we assume mean SFRs of $0.7 M_{\odot}$ yr$^{-1}$ for the faint and the luminous samples. This yields $L_{\text{HMXB}}$ of $1.4 \times 10^{39}$ erg s$^{-1}$, $1.4 \times 10^{38}$ erg s$^{-1}$, $9.8 \times 10^{38}$ erg s$^{-1}$, and $9.8 \times 10^{38}$ erg s$^{-1}$, respectively. Therefore the average total XRB luminosity for the late-type, early-type, faint, and luminous galaxies, respectively, is predicted to be $2.7\times10^{39}$ erg s$^{-1}$, $2.4\times10^{39}$ erg s$^{-1}$, $2.0\times10^{39}$ erg s$^{-1}$, and $5.2\times10^{39}$ erg s$^{-1}$. 

Comparing these predictions to the median total luminosities within 50 kpc (Figure 16), we see that XRBs are predicted to comprise 43\%, 17\%, 41\%, and 18\% of the total emission for the late-type, early-type, faint, and luminous galaxies, respectively. These fractions are lower than the $59\%$ fraction inferred from the spectral hardness ratios, but  given the huge uncertainties in each method, the rough agreement is still encouraging. Our (conservative) conclusion is that the data seem to suggest about half of the emission belongs in the point source component, and about 2/3 of the remaining (extended) emission is due to X-ray binaries, but there is a lot of room for variations between different samples and for complications in the spectral modeling. In particular, the luminous and the early-type galaxies seem to have a higher fraction of hot gas emission than the other samples.

\subsection{The $L_X$-$L_K$ relation}

Now that we have X-ray luminosities for the hot gas around our samples of isolated galaxies, we present the $L_X$-$L_K$ relation for our samples. Figure 19 shows an $L_X$-$L_K$ plot with our three samples with the strongest suggestion of extended emission -  the luminous galaxies and the early-type and late-type luminous subsamples. We also inlcude data from other recent measurements (in the same 0.5--2.0 keV energy band) of isolated elliptical galaxies (\citealt{Mulchaey2010}, \citealt{Memola2009}). For our samples, we plot the mean exposure-weighted $L_K$ (with errorbars enclosing 68\% of the galaxies in each sample) and the median of the pdf of the X-ray luminosity of the extended component out to 50 kpc (along with the 68\% confidence interval on this luminosity). We also indicate with an X symbol the median luminosity after subtracting the X-ray binary luminosity as estimated from the scaling relations (section 7.3). This is probably an over-estimate of the effect of XRBs, however, since $\sim 50\%$ of the X-ray binary emission may be in the point source component instead of the extended component. Our 2MIG luminous galaxies seem to fall within the scatter of the $L_X$-$L_K$ relation, especially after subtracting the estimated XRB emission. 

\begin{figure}
\plotone{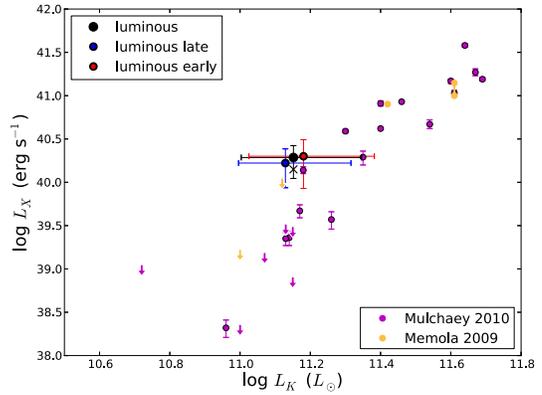}
\caption{  $L_X$-$L_K$ relation for all the luminous, the luminous late-type, and the luminous early-type galaxies. The value of $L_K$ is the mean exposure time-weighted value for each sample, and the error bars enclose the central 68\% of the galaxies in each sample. The value of $L_X$ is the median X-ray luminosity of the extended component of the emission, integrated out to 50 kpc, and the error bars enclose the 68\% confidence interval on this luminosity. The X symbols show the estimated $L_X$ after subtracting the maximum estimated emission from X-ray binaries. For comparison we also include data from other studies of isolated elliptical galaxies. }
\end{figure}

\subsection{Hot Gas Mass}

Finally, after analyzing the X-ray luminosities of the hot gas, we turn to the hot gas mass. We estimate the hot gas mass simply by integrating the $\beta$-model and marginalizing over $\beta$, $r_0$, and $A_{\beta}$. Unfortunately, the measured pdf for $\beta$ does not provide a reliable guide to the true value of $\beta$ (see Figure 7). Moreover, we do not know the outer extent of the hot gas since the detected emission drops off beyond about 50 kpc. Therefore we restrict the analysis of the gas mass to within a radius of 50 kpc, where the gas luminosity is fairly well constrained and variations in $\beta$ are not yet too significant (the slope becomes increasingly important for the mass at larger radii). 

We also need a way to relate $\beta$, $r_0$, and $A_{\beta}$ to the central density $n_0$. We can connect these parameters using the spectral model (section 7.1). For an APEC model with $kT = 0.2$ keV, $Z = 0.3 Z_{\odot}$, and $N_H = 5\times10^{20}$ cm$^{-2}$, we simulated observations with Webspec\footnote{http://heasarc.gsfc.nasa.gov/webspec/webspec.html} and found that the observed 0.5-2.0 keV count rate is equal to 19.32 times the normalization of the APEC model. Unfortunately, at this temperature the proportionality constant is nearly linearly dependent on the metallicity of the gas, and this metallicity could plausibly be anywhere from $0.1 Z_{\odot}$ to $1.0 Z_{\odot}$. This metallicity uncertainty introduces a factor of three uncertainty in the estimate of the normalization.  Even worse, the gas could even be incompletely mixed, or host a metallicity gradient, and we have no way to account for these possibilities, so the normalization is potentially even more uncertain. 

After estimating the normalization of the APEC model from the count rate, we can relate the normalization to the central density using the relation

\begin{equation}\text{norm} \equiv \frac{10^{-14}}{4 \pi d_L} \int{n_e n_H dV}\end{equation}

where $d_L$ is the luminosity distance to the galaxy, and $n_e$ and $n_H$ and the electron and Hydrogen column densities, respectively. Substituting the $\beta$-profile into the equations for $n_e$ and $n_H$, converting $n_H$ into $n_e$, and setting $d_L = 100$ Mpc, we can use this equation to compute the pdf for $n_0$ for each sample, and then integrate the mass profile out to 50 kpc to get a gas mass pdf. 

To account for contamination from X-ray binaries, we divide the values of $A_{\beta}$ from the MCMC fit by 3, since we estimate above (section 7.3) that 2/3 of the extended emission comes from X-ray binaries. This implicitly assumes the X-ray binaries have the same density profile as the hot gas, which is probably not correct, but with the data we have this is the best we can do. The resulting pdfs for the gas mass are shown in Figure 20.

\begin{figure}
\plotone{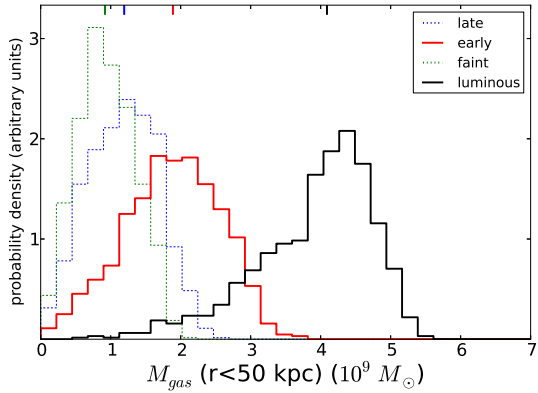}
\caption{  Probability distribution functions (pdfs) for the hot gas mass within 50 kpc of the galaxy. The median of each distribution is marked with a vertical dash at the top of the plot. The masses were corrected downwards to account for contamination from X-ray binaries. The pdfs for the late-type and faint subsamples are shown as dashed lines because the statistical significance of the extended component well below 90\%. The gas was assumed to have $kT = 0.2$ keV and $Z = 0.3 Z_{\odot}$; uncertainties in the metallicity of the gas lead to additional uncertainties in the inferred mass on the order of 75\%.}
\end{figure}

\subsubsection{Accretion Rates}

Finally, we can combine our knowledge of the mass profiles and the luminosity profiles to estimate the cooling rate of the hot gas. The total amount of cooling is set by the X-ray luminosity of the gas at small radii, and we have some confidence in our ability to measure this quantity. In the absence of nongravitational heating, the cooling rate is equal to the hot-mode accretion rate onto the central galaxy. The accreted gas can then serve as fuel for star formation in the galaxy. 

However, at least for early-type galaxies, the cooling rates measured in pointed observations ($\sim 1 M_{\odot}$ yr$^{-1}$, e.g. \citealt{Nulsen1984}, \citealt{Fabbiano1989}) are typically much larger than the (very low) star formation rates typically measured in these galaxies (e.g. \citealt{OConnell1999}). This is the so-called ``cooling-flow'' problem in elliptical galaxies. It is thought that the X-ray cooling is counteracted by heating from AGN feedback or other heating processes (\citealt{Rosner1989}, \citealt{Tucker1997}, \citealt{Mathews2003}), so that the actual hot-mode accretion rate is closer to the star formation rate than the cooling rate.

On the other hand, it obviously should not be possible for the hot mode accretion rate to be higher than the X-ray cooling rate, so the cooling rate sets an upper bound on the hot mode accretion rate (although other modes of accretion can still be extremely important). Moreover, we are unlikely to underestimate significantly the average cooling rate, since our simulations showed that we can recover the luminosity to 30\% or better, and that if the true luminosity were higher we could recover it even more accurately. Since the cooling rate is set by the behavior of the gas at small radii, this quantity is also not very sensitive to the slope of the density profile.

We compute the cooling rate by defining a cooling radius, estimating the thermal energy of the hot gas within the cooling radius, and dividing by the X-ray luminosity to infer a cooling time. The cooling rate is then the gas mass within the cooling radius, divided by the cooling time. Following \citet{Fukugita2006}, we start with the cooling time:

\begin{equation}\tau(r) = \frac{1.5 n k T}{\Lambda n_e (n - n_e)} \approx \frac{1.5 kT \times 1.92}{\Lambda n_e \times 0.92}\end{equation}

where the latter expression assumes the total particle density $n = 1.92 n_e$. For $T = 10^{6.35}$ K and $Z = 0.3 Z_{\odot}$, $\Lambda = 10^{-22.54}$ erg cm$^3$ s$^{-1}$  \citep{Sutherland1993}. Thus for $\tau = 10$ Gyr, the cooling radius occurs at $n_e = 1.1\times10^{-4}$ cm$^{-3}$. Marginalizing over our density profiles, the median cooling radii for the luminous galaxies, the luminous late-type galaxies, and the luminous early-type galaxies are 54 kpc, 48 kpc, and 46 kpc, respectively. The median implied cooling rates are $0.4 M_{\odot}$ yr$^{-1}$, $0.3 M_{\odot}$ yr$^{-1}$, and $0.4 M_{\odot}$ yr$^{-1}$, respectively. 

The inferred cooling rate around our luminous early-type galaxies is consistent with cooling rates measured around ellipticals in pointed observations, and illustrates the same cooling-flow problem, as noted above. On the other hand, for the luminous late-type galaxies, our inferred accretion rate is a few times lower than the expected star formation rate. A similar situation entails for the Milky Way, which also has a hot gas accretion rate several times smaller than the observed star formation rate (\citealt{Anderson2010}, \citealt{Putman2012}), and the implied accretion rates from the hot coronae around the \citet{Tullmann2006} sample of star-forming spirals are also lower than the observed star formation rates. Therefore, the conclusion must be that hot mode accretion on its own is insufficient to fuel star formation in isolated massive spirals. Either additional modes of accretion must supplement the hot gas cooling, or mass lost from stellar winds \citep{Leitner2011} must make up the difference.

\subsubsection{Baryon budgets}

As mentioned above, in our model the total hot gas is primarily determined by the value of $\beta$ and by the outer radius of the hot halo. We do not have strong constraints on either of these parameters, other than that the hot halo seems to extend out to at least 50 kpc, so we do not integrate the mass profiles beyond 50 kpc. In this section, however, we speculate briefly on the possible behavior at larger radii. 

Based on previous observations described in section 7, we believe the likeliest value for $\beta$ is $\beta \approx 0.5$, but the flattest density profiles in simulations have $\beta \approx 0.35$ (e.g. \citealt{Kaufmann2009}) and the steepest observed profiles approach $\beta \approx 0.7$ (numerical predictions by \citealt{Feldmann2012} also predict a similarly large value of $\beta$), thus motivating the range of values we explored in our simulated images. If we integrate the $\beta$-model from 50 kpc outwards, we can get a sense of the potential hot gas mass at larger radii. Figure 21 shows the ratio of the hot gas mass out to 200 or 300 kpc to the hot gas mass within 50 kpc, as a function of $\beta$. The range on each distribution reflects the effect of varying the core radius $r_0$ between 0.5 kpc and 5.0 kpc -- the effect of varying $r_0$ is not very significant. Within 200 kpc, the hot gas mass can increase from a factor of a few up to a factor of 15, and if we extend the profile to 300 kpc the total mass can increase very significantly for the flatter profiles. 

\begin{figure}
\plotone{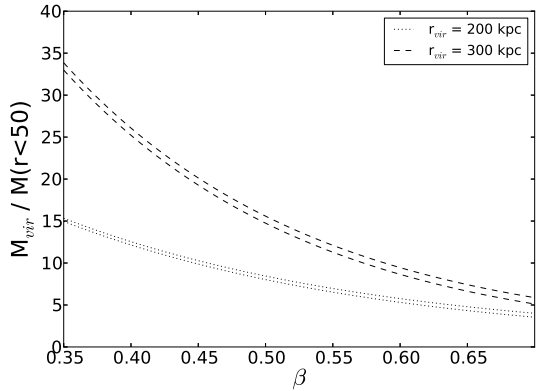}
\caption{  Ratio of the hot gas mass within 200 or 300 kpc to the hot gas mass within 50 kpc, assuming the hot gas follows a $\beta$-model. The range on each distribution reflects the effect of varying the core radius $r_0$ between 0.5 kpc and 5.0 kpc. }
\end{figure}

From section 7.5, the median masses of the hot gas within 50 kpc around the early-type galaxies and the luminous galaxies are $1.9\times10^9 M_{\odot}$ and $4.1\times10^9 M_{\odot}$. For $\beta = 0.5$, integrating out to 200 kpc would increase these masses to $1.5\times10^{10} M_{\odot}$ and $3.3\times10^{10} M_{\odot}$, respectively. These inferred gas masses are much larger than the observed masses, and are now comparable to the average exposure-weighted stellar masses of the galaxies in their respective samples. Since the stars in the average massive galaxy contain less than half of the baryons expected in the halo based on the cosmological baryon fraction (e.g. \citealt{McGaugh2010}, the hot halo appears to be a tempting possible reservoir for these missing baryons. The slope of the density profile would have to be somewhat flatter than $0.5$ to account for all the missing baryons, but the \emph{ROSAT} observations do not constrain this extrapolation, and indeed it is very difficult to place any constraints on X-ray emission at these radii. However, as we pointed out in \citet{Anderson2010}, massive hot halos extending to such large radii violate X-ray absorption-line constraints (and decreasing $\beta$ exacerbates the problem). Therefore we think the hot halo probably does not follow a $\beta$-model out to hundreds of kpc.

\subsection{Comparison with previous work}

We briefly compare our results to results from X-ray stacking analyses of other samples of galaxies. Often, this sort of stacking has been performed for high-redshift galaxies in the deep fields (e.g. $z\sim3$ Lyman-break galaxies; \citealt{Nandra2002}, \citealt{Lehmer2005}, and \citealt{Laird2006}; $z\sim1$ UV-luminous galaxies; \citealt{Laird2005}) which have very different properties from the nearby isolated galaxies in our sample. A few papers do examine ``normal'' galaxies at lower redshift, however. \citet{Lehmer2007} stacked 539 early-type galaxies from $0.1\lapprox z \lapprox0.7$, and found mean $z \approx 0.25$ X-ray luminosities (0.5-2.0 keV) for their luminous and faint samples of $10^{40.2\pm0.2}$ erg s$^{-1}$ and $10^{39.6\pm0.4}$ erg s$^{-1}$, respectively, with no evidence for redshift evolution. While they use a different optical classification for luminous and faint galaxies, and a smaller aperture than our analysis (based on maximizing the observed signal-to-noise), their results are consistent with ours. \citet{Watson2009} perform a similar analysis on 6146 galaxies in the AGN and Galaxy Evolution Survey (AGES) over $0.1 \lapprox z \lapprox 0.5$. They parameterize their results in terms of $L_K$ and $(1+z)$; plugging in our mean values for these quantities, we infer a prediction of $L_X = 2.4\pm0.7 \times 10^{40}$ erg s$^{-1}$, which is not too different from our value of $L_X = 1.0\pm0.3 \times 10^{40}$ erg s$^{-1}$. Their values could also be slightly higher because their sample is not restricted to isolated galaxies; galaxies in denser environments might be expected to have higher rates of SMBH accretion, and also they have additional emission from a hot intracluster/intragroup medium.

\section{Conclusions}

We examine 2496 nearby isolated galaxies from the 2MASS  Isolated Galaxies survey, and stack \emph{ROSAT} All-Sky Survey images of 2165 of these galaxies. We also divide the full 2165 galaxies by morphology (into subsamples of late-type and early-type galaxies) and luminosity (into subsamples of luminous and faint galaxies), and examine each subsample in turn. We also examine further subsamples of the luminous galaxies, divided into luminous early-type and luminous late-type galaxies, and we produce a stacked image of random positions on the sky as well as a null sample. For each sample, we stacked all the 0.5-2.0 keV photons within 500 projected kpc of the galaxy, and also stacked the RASS exposure map out to the same radius, weighing the exposure map by the inverse square of the distance to the galaxy. 

In the stacked images, we model the surface brightness profiles as the sum of a uniform background, a point source convolved with an empirically determined PSF, and an extended component (a $\beta$-model) convolved with the same PSF. We also include in the model an ``effective vignetting'' caused by incomplete coverage at larger radii in some images, which we measured from the stacked effective exposure map. We employ an MCMC analysis to compare the model to the observed surface brightness profile. 

For each of our samples, except for the null sample, we also create a variety of simulated images which appear similar to the real images, and test how well the MCMC analysis can recover the various image parameters. In general the MCMC analysis can recover the number of source counts within 50 kpc of the central galaxy, but it cannot recover the slope and core radius of the extended component reliably. It also has mixed success in distinguishing emission in the point source component and the extended component, but the MCMC analysis, a likelihood ratio test, and the hardness ratio of the emission all suggest the emission is composed of a mixture of the two components in fairly comparable amounts.

We examine the Poisson statistics of the counts within 50 projected kpc of the center of each image in order to estimate the statistical significance of detecting emission in each stacked image. We detect emission in all stacked images, except the null image, at $>99\%$ confidence. We then employ likelihood ratio tests and simulate images in order to estimate the statistical significance of detecting {\it extended} (beyond 5 kpc) emission in each stacked image. For five out of seven samples (all galaxies, early-type galaxies, luminous galaxies, luminous late-type galaxies, and luminous early-type galaxies) an extended component has a statistical significance of $\gapprox 90\%$ (and the luminous galaxies have an extended component at $>99\%$); the other two samples have extended emission at much lower significance. We compute the average X-ray luminosity within 50 kpc of each sample of galaxies, and we compute the average X-ray luminosity of the extended component within 50 kpc as well, with the caveat that the extended component may not be statistically necessary in some cases. 

From the observed hardness ratios of the emission, and from empirical scaling relations, we estimate the contribution of X-ray binaries to the luminosity within 50 kpc of the galaxy in each of our samples. While the uncertainties are very large, both methods generally agree, and suggest that about 2/3 of the extended emission is due to X-ray binaries, with the remaining 1/3 of the extended emission attributed to hot gas. We also estimate that the point source component is roughly evenly distributed between a weakly accreting SMBH and X-ray binaries in the central 5 kpc of the galaxy. We can then infer the mass of the hot gas within 50 kpc, again with the caveat that the extended component may not be statistically required for some of the stacked samples of galaxies. 

We find that the average X-ray luminosity of the early-type galaxies is about twice the X-ray luminosity of the late-type galaxies, and that the statistical significance of the extended emission around the early-type galaxies is also much higher. The early-type galaxies therefore seem to have larger hot gas halos than the late-type galaxies. It is unclear, however, if this is related to the morphology of the galaxies or simply a consequence of the early-type galaxies having larger average stellar masses and luminosities than the late-type galaxies. For the luminous subsamples, the late-type galaxies and early-type galaxies have the same X-ray luminosities and hot gas masses, within the (large) uncertainties, although the luminous late-type galaxies have extended emission with higher statistical confidence. 

Finally, for the samples where we measure a hot gas mass, we also infer a hot gas accretion rate onto the central galaxy. These accretion rates are a few times smaller than the expected star formation rates in the luminous late-type galaxies, and a few times larger than the expected star formation rates in the luminous early-type galaxies. We comment briefly on the expected hot gas masses if the $\beta$-model is extrapolated out to the virial radius. If the hot gas has a typical value of $\beta \approx 0.5$, and extends to the virial radius, the total hot gas mass would be expected to be 10-20 times larger than the value we measure. Flatter values of $\beta$ would increase this mass significantly, as would abundance gradients in the hot halo, but at present we have no observational constraint on either of these possibilities. 

All of the measured and derived quantities discussed above are presented in Table 4. Median values of the pdf are listed, along with 68\% confidence intervals (identified as the statistical errors). We also estimate the size of the systematic uncertainties on each quantity and list those uncertainties as well; for detailed discussion of the systematic uncertainties, see the appropriate sections of the paper. 

In the next few years, the E-Rosita mission is expected to launch, and to produce another all-sky survey with coverage in the soft X-rays. This survey could be up to an order of magnitude deeper than the RASS. It will be very interesting to examine the properties of stacked galaxies in this survey, as we should be able to characterize the extended emission with much more precision.

\section{Acknowledgements}
The authors would like to thank E. Bell, E. Gallo, O. Gnedin, B. Miller, C. Miller, J. Miller, M. Ruszkowski, C. Slater, R. Smith, and A. Vikhlinin for very helpful conversations and suggestions which greatly improved this manuscript. We would also like to thank the referee for suggestions and references that greatly improved the result. This research has made use of data and/or software provided by the High Energy Astrophysics Science Archive Research Center (HEASARC), which is a service of the Astrophysics Science Division at NASA/GSFC and the High Energy Astrophysics Division of the Smithsonian Astrophysical Observatory. This research has made use of NASA's Astrophysics Data System. M.E.A. also gratefully acknowledges support from the NSF in the form of a Graduate Research Fellowship.

\begin{sidewaystable*}
\begin{tiny}
\caption{Summary of measurements and derived quantities}
\begin{tabular}{ccccccccc}
\hline
\hline
sample & $L_K$ & $1-p$(any) & $1-p$(ext) & $ L_{X,<50}$ & $L_{X\text{,ext,}<50}$ &$L_{X,<300}$ & $M_{\text{gas},<50}$ & ${\dot M}$\\ 
 & $(10^{10} L_{\odot})$ &    &     & $(10^{39}$ erg s$^{-1}$) & $(10^{39}$ erg s$^{-1}$)  & $(10^{39}$ erg s$^{-1}$) & $(10^9 M_{\odot})$ & $(M_{\odot}$ yr$^{-1})$ \\
 \hline
 
All galaxies &  $5.1^{+10.5}_{-2.2}$& $4.4\times10^{-16}$ $(8.1 \sigma)$ & 0.073 &$10.0\pm0.8$(stat) $\pm2.0$(sys) &$5.5^{+3.1}_{-3.4}$(stat)$\pm1.1$(sys)&$10.6^{+0.9}_{_0.8}$(stat) $\pm2.1$(sys) &$1.6_{-0.6}^{+0.5}$(stat)$\pm{1.2}$(sys)&$0.09\pm0.06$\\
Late-type galaxies & $4.4^{+9.7}_{-1.9}$ &$0.000446$  & 0.347& $6.3\pm0.9$(stat) $ \pm0.9$(sys)& $2.9^{+2.5}_{-2.0}$(stat)$\pm0.4$(sys)* &$6.7^{+1.1}_{-0.9}${(stat)} $\pm1.0$(sys)&$1.2^{+0.5}_{-0.6}$(stat)$ \pm0.8$(sys)*&$0.05\pm0.04$*\\
Early-type galaxies & $7.4^{+12.0}_{-3.1}$ &$3.0\times10^{-9}$ $(5.9\sigma)$ & 0.107 & $14.2^{+1.6}_{-1.8}${(stat)} $\pm4.3$(sys)&$7.7^{+4.6}_{-4.8}${{(stat)}} $\pm2.3${(sys)}&$15.0^{1.8}_{-1.7}${(stat)} $\pm4.5$(sys)&$1.9\pm0.7${(stat)} $\pm1.3$(sys)&$0.1\pm0.1$ \\
Faint galaxies & $3.3^{+4.0}_{-1.2}$ & 0.005312 & 0.528 & $4.9^{+0.7}_{-0.8}${(stat)} $\pm1.5$(sys)&$2.1_{-1.6}^{+2.0}$(stat)$\pm0.6${(sys)}*&$5.2\pm0.8${(stat)} $\pm1.6$(sys)&$0.9^{+0.5}_{-0.4}$(stat)$\pm0.6$(sys)*&$0.04\pm0.03$*\\
Luminous galaxies & $14.2^{+8.0}_{-4.1}$ &$2.6\times10^{-23}$ $(9.9\sigma)$& 0.007&$28.3^{+3.9}_{-4.1}${(stat)} $\pm4.2$(sys)&$19.3^{+7.2}_{-8.2}$(stat) $\pm2.9${(sys)}&$36.2^{+4.5}_{-4.0}${(stat)} $\pm5.4$(sys)&$4.1^{+0.6}_{-1.0}${(stat)} $\pm2.9$(sys)&$0.4^{+0.1}_{-0.2}$\\
Luminous late-type galaxies & $13.5^{+7.2}_{-3.6}$ &  $0.000025$ &0.013&$25.9^{+4.6}_{-4.3}${(stat)} $\pm3.9$(sys)&$16.7^{+7.6}_{-8.0}$(stat) $\pm2.5$(sys)&$31.5^{+5.0}_{-4.4}$(stat) $\pm4.7$(sys)&$3.6^{+0.8}_{-1.1}$(stat) $\pm2.5$(sys) &$0.3^{+0.1}_{-0.2}$\\
Luminous early-type galaxies & $15.2^{+9.0}_{-4.6}$ & $3.2\times10^{-15}$ $(7.9\sigma)$ & $0.079$&$35.6^{+5.6}_{-5.2}$(stat) $\pm5.3$(sys)&$20.0^{+11.0}_{-11.5}$(stat) $\pm3.0$(sys)&$41.0^{+6.0}_{-5.1}$(stat) $\pm6.1$(sys)&$3.5_{-1.4}^{+1.2}$(stat) $\pm 2.5$(sys) &$0.4\pm0.2$\\
\hline
\end{tabular}
\end{tiny}
\vspace{0.2cm}
{\center Measured and derived quantities for stacks of various samples of 2MIG galaxies. $L_K$ is the mean exposure-weighted K-band luminosity of a galaxy in each sample, and the range enclosing the central 68\% of luminosities is indicated. The next two columns show the statistical significance of detections of any emission and of extended emission within the central 50 projected kpc of the stacked images. $L_{X,<50}$ is the median of the pdf of the 0.5-2.0 keV X-ray luminosity within 50 physical kpc of the galaxy in each stack. The 68\% confidence interval is indicated with (stat) and the estimated impact of systematic errors is indicated with (sys). $L_{X,\text{ext},<50}$ shows the luminosity of extended emission within 50 kpc, in the same format.  $L_{X,<300}$ is the same as $L_{X,<50}$ except the emission is integrated out to 300 kpc from the galaxy instead of 50 kpc. $M_{\text{gas},<50}$ is the estimated hot gas mass within 50 kpc of the galaxy, after correcting for X-ray binary contamination. The systematic uncertainty accounts for metallicities ranging from $0.1 Z_{\odot}$ to $1.0 Z_{\odot}$. Finally, ${\dot M}$ is the estimated accretion rate of hot gas onto the central galaxy (and 68\% confidence intervals on the median value). Asterisks denote the quantities based on extended emission detected at $\lapprox 90\%$ significance.  }
\end{sidewaystable*}
\clearpage

\bibliographystyle{apj}
\bibliography{paper.bib}{}

\end{document}